\documentclass[acmlarge]{acmart}
\usepackage{booktabs} 
\usepackage[ruled]{algorithm2e} 
\usepackage{amsmath} 
\usepackage{siunitx} 
\usepackage{graphicx, multirow}
\usepackage{subcaption}
\usepackage{tabularx}

\usepackage[obeyFinal]{easy-todo} 
\usepackage{capt-of}
\usepackage[normalem]{ulem}
\usepackage[export]{adjustbox} 
\usepackage{tikz} 

\SetAlFnt{\small}
\SetAlCapFnt{\small}
\SetAlCapNameFnt{\small}
\SetAlCapHSkip{0pt}
\IncMargin{-\parindent}

\acmJournal{JOCCH}
\acmVolume{12}
\acmNumber{4}
\acmArticle{0}
\acmYear{2019}
\acmMonth{12}
\acmArticleSeq{01}

\setcopyright{acmlicensed}

\acmDOI{0000001.0000001}

\received{May 2018}
\received{October 2018}
\received{January 2019}
\received[accepted]{March 2019}

\begin{document}
\title{Gloss, Color and Topography Scanning for Reproducing a Painting's Appearance using 3D printing} 

\author{Willemijn Elkhuizen}
\orcid{0000-0002-2223-9114}
\email{w.s.elkhuizen@tudelft.nl}
\author{Tessa Essers}
\email{t.t.w.essers@tudelft.nl}
\author{Yu Song}
\email{y.song@tudelft.nl}
\orcid{0000-0002-9542-1312}
\author{Jo Geraedts}
\email{j.m.p.geraedts@tudelft.nl}
\affiliation{%
  \institution{Design Engineering, Delft University of Technology}
  \streetaddress{Landbergstraat 15}
  \city{Delft}
  \postcode{2628CE}
  \country{The Netherlands}} 
\author{Clemens Weijkamp}
\email{clemens.weijkamp@oce.com}
\affiliation{%
  \institution{Oc\'{e} Technologies BV}
  \streetaddress{Sint Urbanusweg 43}
  \city{Venlo}
  \postcode{5914CA}
  \country{The Netherlands}}
\author{Joris Dik}
\email{j.dik@tudelft.nl}
\affiliation{%
  \institution{Materials Science and Engineering, Delft University of Technology}
  \streetaddress{Mekelweg 2}
  \postcode{2628CD}
  \city{Delft}
  \country{The Netherlands}}
\author{Sylvia Pont}  
\email{s.c.pont@tudelft.nl}
\affiliation{%
  \institution{Industrial Design, Delft University of Technology}
  \streetaddress{Landbergstraat 15}
  \city{Delft}
  \postcode{2628CE}
  \country{The Netherlands}}

\begin{abstract}
High fidelity reproductions of paintings provide new opportunities to museums in preserving and providing access to cultural heritage. This paper presents an integrated system which is able to capture and fabricate color, topography and gloss of a painting, of which gloss capturing forms the most important contribution. A 3D imaging system, utilizing stereo imaging combined with fringe projection, is extended to capture spatially-varying gloss, based on the effect of specular reflectance polarization. The gloss is measured by sampling the specular reflection around Brewster's angle, where these reflections are effectively polarized, and can be separated from the unpolarized, diffuse reflectance. Off-center gloss measurements are calibrated relative to the center measurement. Off-specular gloss measurements, following from local variation of the surface normal, are masked based on the height map and corrected. Shadowed regions, caused by the 3D relief, are treated similarly. The area of a single capture is approximately \SI{180x90}{\milli\meter} at a resolution of \SI{25x25}{\micro\meter}. Aligned color, height, and gloss tiles are stitched together off-line, registering overlapping color regions. The resulting color, height and gloss maps are inputs for the poly-jet 3D printer. Two paintings were reproduced to verify the effectiveness and efficiency of the proposed system. One painting was scanned four times, consecutively rotated by \SI{90}{\degree}, to evaluate the influence of the scanning system geometric configuration on the gloss measurement. Experimental results show that the method is sufficiently fast for practical application, i.e. to scan a whole painting within eight hours, during closing hours of a museum. The results can well be used for the purpose of physical reproduction and other applications needing first-order estimates of the appearance (e.g. conservation diagnostics and condition reports). Our method to extend appearance scanning with gloss measurements is a valuable addition in the quest for realistic reproductions, in terms of its practical applicability - number of images needed for reconstruction and speed - and its perceptual added value, when added to color and topography reproduction. 
\end{abstract}

%
%


%
%


\thanks{
This article is a revised and extended version of Elkhuizen et al. \cite{Elkhuizen2017} published in GCH 2017 by Eurographics Proceedings \copyright, The Eurographics Association 2017. Reproduced by kind permission of the Eurographics Association. Authors' addresses: W.S. Elkhuizen (corresponding author), T.T.W. Essers, Y. Song, J.M.P. Geraedts, S.C. Pont, Delft University of Technology, Landbergstraat 15, 2628 CE  Delft, The Netherlands; email: w.s.elkhuizen@tudelft.nl; C. Weijkamp, Oc\'e Technologies BV, St. Urbanusweg 43, 5914 CA Venlo, The Netherlands; J. Dik, Delft University of Technology, Mekelweg 2, 2628 CD  Delft, The Netherlands.
}

\maketitle

\renewcommand{\shortauthors}{W.S. Elkhuizen et al.}

\section{Introduction} \label{sec:Introduction}
Developments in 3D scanning and 3D printing systems provide new opportunities to create high-fidelity physical reproductions of paintings. Facsimiles (one-to-one reproductions simulating the artifact's appearance), are already made of  artifacts like manuscripts, for instance to support storytelling in exhibitions, when the original is too fragile to show. Up to recently, copies with such likeness did not exist for paintings \cite{Brigstocke2001}. Facsimiles of paintings, in addition to the original artwork, can play a role in museums' missions of preserving as well as providing access to cultural heritage. Possible applications of scanning and reproduction include: Multi-modal documentation of an artwork (e.g. adding to high-resolution photography, infrared, and X-ray imaging, as shown in the Bosch Research and Conservation Project \cite{Erdmann2016, Ilsink2016}), showing an artwork outside a museum context, showing reconstructions of the original state of an artwork  (e.g. \cite{Elkhuizen2016}), creating records of an artwork in different stages of a restoration process, or selling high-end reproductions. 

All these applications require reproductions that closely resemble the original artwork's appearance. For this, various modalities need to be captured and reproduced: color, topography (three dimensional height variations of the surface), gloss, and translucency. Zaman et al. presented a system which is able to reproduce only the color and topography of a painting's surface using 3D scanning and 3D printing technology \cite{Zaman2014}. Using this system, three reproductions were made and compared to the original painting. It was found that the uniform gloss of the reproductions' surfaces, made these look artificial or even "plastic", which clearly distinguishes them from paintings, which exhibit spatially-varying gloss \cite{Elkhuizen2014}.

In this paper, we present an integrated system which is able to capture the color, topography and gloss of paintings for the purpose of 3D printing, of which gloss \emph{capturing} forms our most important contribution. The main contributions of this paper are: a) a novel approach to gloss capturing using reflection polarization, and b) an integrated scanning procedure which is sufficiently fast for appearance capturing of large areas for practical application. Two oil paintings, named \textit{Two Wrestling Figures} and \textit{Sunflowers}, both painted \emph{in the style of} Vincent Van Gogh, are reproduced. The painting \textit{Two Wrestling Figures} is a reconstruction of a lost painting by Vincent Van Gogh, which was rediscovered using XRF scanning \cite{Alfeld2013a}, and recreated, using oil paint, for a Dutch television program \cite{Pos2017}. The \textit{Sunflowers} painting was made by an anonymous painter.

The remainder of this paper is arranged as follows: in section \ref{sec:Related work}, literature on capturing and reproduction of material appearance is reviewed. Section \ref{sec:Materials} presents the system and and section \ref{sec:Method} the approach that is deployed to measure and fabricate color, topography and spatially-varying gloss. An experiment was conducted utilizing the proposed system and the scanning and printed results are presented in section \ref{sec:Results}. The advantages and limitations of the proposed system are discussed in section \ref{sec:Discussion} and conclusions are drawn in section \ref{sec:Conclusions}. 

\section{Related work} \label{sec:Related work}%
This section reviews the state-of-the-art regarding the reproduction of material appearance, covering capturing as well as fabrication of material appearance. The focus of the review lies on methods targeted at capturing and/or fabricating the appearance of planar (but non-flat) surfaces like paintings, bas-reliefs, parchments, and fabrics, which exhibit spatially-varying reflectance. 

\subsection{2D color reproduction} 
2D color reproduction is a mature field, where standards and guidelines exist to support the color reproduction workflow from capturing an image (e.g. for digitizing cultural heritage \cite{FADGI2010, Dormolen2012-EN}) to the conversion of this data for printing  \cite{ISO2015}. Limitations of RGB imaging and CMYK printing are known; for instance in terms of a limited printer gamut, and color mismatch in different illumination conditions (metamerism). Multi-spectral imaging systems (summarized in \cite{Fischer2006}) have been developed to capture the diffuse reflectance of paintings more accurately. Berns et al.\cite{Berns1998, Berns2008} combined multi-spectral imaging with multichannel printing to create painting reproductions, minimizing color metamerism effects. Although successful, they speculated that this improvement alone is probably too small to justify the investment in developing a multi-spectral reproduction workflow \cite{Berns1998}. Furthermore, even if the color would be very accurately reproduced for all illumination conditions, this does not comprise the total appearance. Our approach relies on RGB imaging and CMYK (and White) printing, for replicating the diffuse color appearance. 

\subsection{2D gloss capturing} 
Various approaches have been proposed to capture the angular appearance variation of painted surfaces, largely for the purpose of computer rendering \cite{Westlund2002, Chen2007, Tominaga2008, Chen2011}. These approaches assume a spatially uniform reflectance, represented by a Bidirectional Reflection Distribution Function (BRDF) \cite{Nicodemus1977}, and therefore use the angular reflectance measurements of individual points across all points on the surface.
However, in a previous evaluation of 3D printed reproductions \cite{Elkhuizen2014} conservation experts remarked that the lack of \emph{spatial variation} of gloss is one of the aspects that distinguishes the reproductions from paintings, meaning these methods do not suit our goal to reproduce the original artwork's appearance.

Angular-spatial appearance variation is often compactly represented by a Spatially-Varying Bidirectional Reflection Distribution Function (SVBRDF), describing the relation between incoming irradiance and the outgoing (reflected) radiance for every point on a surface. In order to achieve a complete representation of the surface reflectance, it would be necessary to measure this relationship for every point on the surface (defined by an x,y coordinate), for every possible incoming irradiance direction (defined by two angles), and every possible outgoing radiance direction (also defined by two angles). As sampling of this full 6-dimensional space is not feasible for every point on a larger surface, a trade-off is made between acquisition speed and measurement accuracy. 
Several approaches employ sparse sampling using point light sources (e.g. \cite{Padfield2002, Redman2007, Hasegawa2011, Paterson2005}). However, high gloss surfaces are not (well) modeled in these approaches. Either the specular reflectance is not separately modeled (in polynomial fitting), or becomes noisy, leading to crosstalk between parameters. As a solution, surfaces are sorted into material groups, and the angular measurements are combined within each group, and shared across the spatial domain. Other approaches also rely on the assumption that the surface is comprised of a limited number of homogeneous materials, grouping appearance into regions, taking advantage of spatial reflectance sharing (e.g. \cite{Wang2008,Dong2010a}). 
For instance, Holroyd et al. \cite{Holroyd2010} took a clustering approach in their synchronous estimation of geometry and surface reflectance. However, as (old) painted surfaces cannot be segmented into a limited set of distinct, uniform materials - they show gradual as well as sudden changes in specular reflectance due to a mixture of materials and other factors which influence the surface state - these approaches are not suitable for our application.

To achieve a denser angular sampling of the surface reflectance, a linear light source has been employed to recover a SVBRDF \cite{Gardner2003, Ren2011, Chen2014}. Alternatively, an LCD screen has been used to project a series of patterns in the frequency domain \cite{Aittala2013}. Similarly Ghosh et al. \cite{Ghosh2009} utilize an LCD projector to project spherical gradient illumination patterns and use the effect of polarization to recover anisotropic specular roughness. These approaches require a large number of images per sample region (see table \ref{tab:SVBRDF_methods}) to accurately estimate the reflectance (model parameters) at each point. 

In contradiction to rendering applications, printing requires a spatial resolution of at least 300dpi (\SI{85}{\micro\meter}) \cite{FograResearchInsituteforMediaTechnologies2018}. At this spatial resolution the above mentioned methods are very time consuming in terms of data acquisition and processing (i.e. due to their angular resolution), limiting their practical use for reproducing whole paintings, which have dimensions typically in the range of 0.5 to 2 \si{\metre\squared}). Moreover, famous paintings often need to be scanned in a limited time slot. In our approach we can suffice with only two reflectance model parameters: the color, representing the diffuse reflectance, and a gloss parameter, representing the magnitude of the specular reflectance peak. To capture these parameters, we need a negligible angular resolution. 
\begin{table}
\begin{center}
	\caption{High sampling density SVBRDF methods, their scanning area and the number of needed individual scans} \label{tab:SVBRDF_methods}
    \begin{tabular}{llll}
  \toprule 
  Method & Authors & No. captures &Scan area\\ 
  &&
  & size (est.)\\
  \midrule
  Linear Lightsource Reflectometry & Gardner et al.\cite{Gardner2003} & 400 & 20x30cm\\
  Generalized Linear Lightsource Reflectometry & Chen et al. \cite{Chen2014} & 240 & 20x30cm\\
  Pocket Reflectometry & Ren et al. \cite{Ren2011} & 900 & 16x17cm\\
  SVBRDF capturing in the Frequency Domain & Aittala et al. \cite{Aittala2013} & 131 & 15x15cm\\
  Second Order Spherical Gradient Illumination & Ghosh et al. \cite{Ghosh2009} & 9 x "large set" & 5x15cm\\
  \bottomrule
\end{tabular} 
\end{center}
\end{table}

Another approach to describe textured surfaces (exhibiting self-shadowing and shading) is the so called Bi-directional Texture Function (introduced by \cite{Dana1999}). This approach is an image based technique, whereby the mesoscopic effects of the surface on its appearance are included into the measurement. This approach does not construct a parameterized representation of the surface, meaning that the specular and diffuse reflectance, shadowing, shading, etc. are not separately modeled for every point. For this reason and the fact that this method also requires a high sampling density (typically 200 images per scan area) to accurately capture the appearance, this approach is also not suitable for our application.

\subsection{2D gloss fabrication} 
In fabricating spatially-varying gloss, several approaches have been demonstrated; Combining inks with various reflectance properties \cite{Matusik2009}, combining a mono-color 3D printed micro texture with a reflective layer and a (2D) color print \cite{Lan2013}; changing the printer parameters to influence the micro-structure of printed surfaces and thereby the gloss \cite{Baar2014}; or half-toning a transparent ink on top of a color print \cite{Baar2015a, Samadzadegan2015, Elkhuizen2015}. The latter approach is the most viable approach in terms of practical applicability (limited number of inks needed), flexibility (can be manipulated independent of sub-layers) and accuracy (in terms of registration), and is therefore also applied in this paper. 

\subsection{3D color and topography capturing} 
Various systems can capture the color as well as topography of paintings' surfaces; using three-color laser scanning \cite{Blais2007,Arius3D2017}, combining line-laser scanning with color imaging \cite{FactumArte2016}, and fringe projection 3D scanning (e.g. \cite{Akca2007,Breuckmann2011,Karaszewski2013}). A limitation of using RGB laser light for the color capturing, is that the narrow spectral bandwidth leads to poor color rendition. A downside of combining two imaging technologies is the need for image registration and potential misalignment of the color and 3D data. A limitation of the fringe projection system (with only one camera) is that the resolution is limited by the projector (which is typically much lower than a camera sensor).
Although examples indicate that the resolution and accuracy of photogrammetry and other passive shape-from-x methods are sufficient for the application of computer rendering (e.g. \cite{CulturalHeritageImaging2017}), we belief that they are less suited to capture high-resolution 3D details for the purpose of creating 3D printed replicas. In using these methods the lack of measured 3D detail can be 'masked' by texture mapping (i.e. shadows and highlights in the color texture, mask the lack of actual depth information). 
Moreover, 3D printing requires higher resolution data, as well as color information free from highlights and shadows. The approach in this paper utilizes the method of Zaman et al. \cite{Zaman2014}, combining fringe projection with stereo vision, for simultaneous, high-resolution capture of color and topography, which mitigates the above mentioned issues. To our knowledge only a few of these scanners \cite{Arius3D2017,Zaman2014,Elkhuizen2017} have been utilized for capturing input data to create physical, full-color reproductions of paintings, which include topography. 

\subsection{3D color and topography fabrication} 
The above mentioned scanners have been used in conjunction with the Elevated Printing technology by Oc{\'{e}} Technologies \cite{ProjectEigerWebsite2017,VerusArt2017} to create 3D printed reproductions of paintings. Other systems currently capable of making high-quality full color 3D prints are the (adapted) Stratasys Connex3 or J750 \cite{Stratasys2018}, or the custom built Multifab printing system \cite{Sitthi-Amorn2015}. Limitations of these systems are the limited print area (max. \SI{490x390}{\milli\meter}), and the translucency of the inks, leading to blurring of fine details, and the need to digitally compensate for these effects, in turn resulting in a lower effective resolution \cite{Brunton2015, Babaei2017, Elek2017, Elkhuizen2018}. Other approaches to physical appearance reproduction of paintings use a hybrid fabrication technique, combining plaster casting, 2D color printing, and artisan hand painting and varnishing to create life-like reproductions \cite{Zalewski2016,FactumArte2016}. 3D reproductions have also been created by the Van Gogh Museum. From publicly available information it can be deduced that these reproductions are created using a hybrid technique of 2D color imaging, 3D scanning, molding and 2D color printing \cite{VanGoghMuseum2017}. A limitation of this reproduction approach is the need for alignment of the color and topography in the fabrication stage. Further details on for instance scanning method, fabrication procedure, scanning and printing resolution, and ultimately the overall reproduction quality are unknown to the authors.\\

In conclusion, currently there is no digital reproduction workflow integrating all modalities of appearance, namely color, topography, gloss and translucency. In the following section we will present an approach to an integrated digital capture and reproduction of the appearance of paintings, which includes the color, topography and gloss variations across the surface. Note that the current setup is not yet able to capture or replicate the translucency of a painted surface (i.e. found in paintings which are built up in various translucent layers, using glazes). 

\section{Materials} \label{sec:Materials}
\subsection{Scanning system} 
\begin{figure}
\centering
  \includegraphics[width=0.75\linewidth]{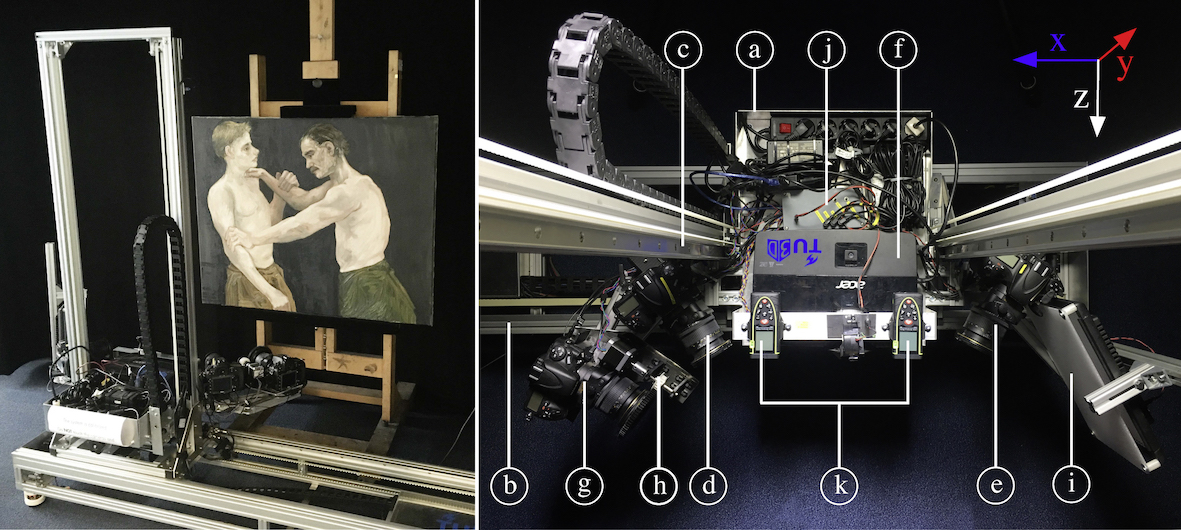}
  \caption{Scanning system. Left: Scanner positioned in front of painting. Right: Top view of (a) the scanning platform guided along (b) a horizontal and (c) vertical frame. 3D scanning module: (d, e) two cameras, and (f) a projector. Gloss scanning module: (g) a camera, (h) a stepper motor driving the rotation of polarization filter, and (i) LED array light source with diffuser. All components are controlled by (j) an Arduino\textsuperscript{\textregistered} micro controller and the scanner is equipped with (k) two distance meters.}
  \label{fig:Scanner}
\end{figure}

The scanning system consists of two modules: the 3D scanning module, which is used to capture color and topography, and the gloss scan module, used for capturing the gloss. The modules capture the same small region of the painting, and their capturing routine runs sequentially for every scan position. Figure \ref{fig:Scanner} shows the scanner and a setup of the experiment. The following paragraphs describe it's components.

\begin{figure}
\centering
  \includegraphics[width=0.5\linewidth]{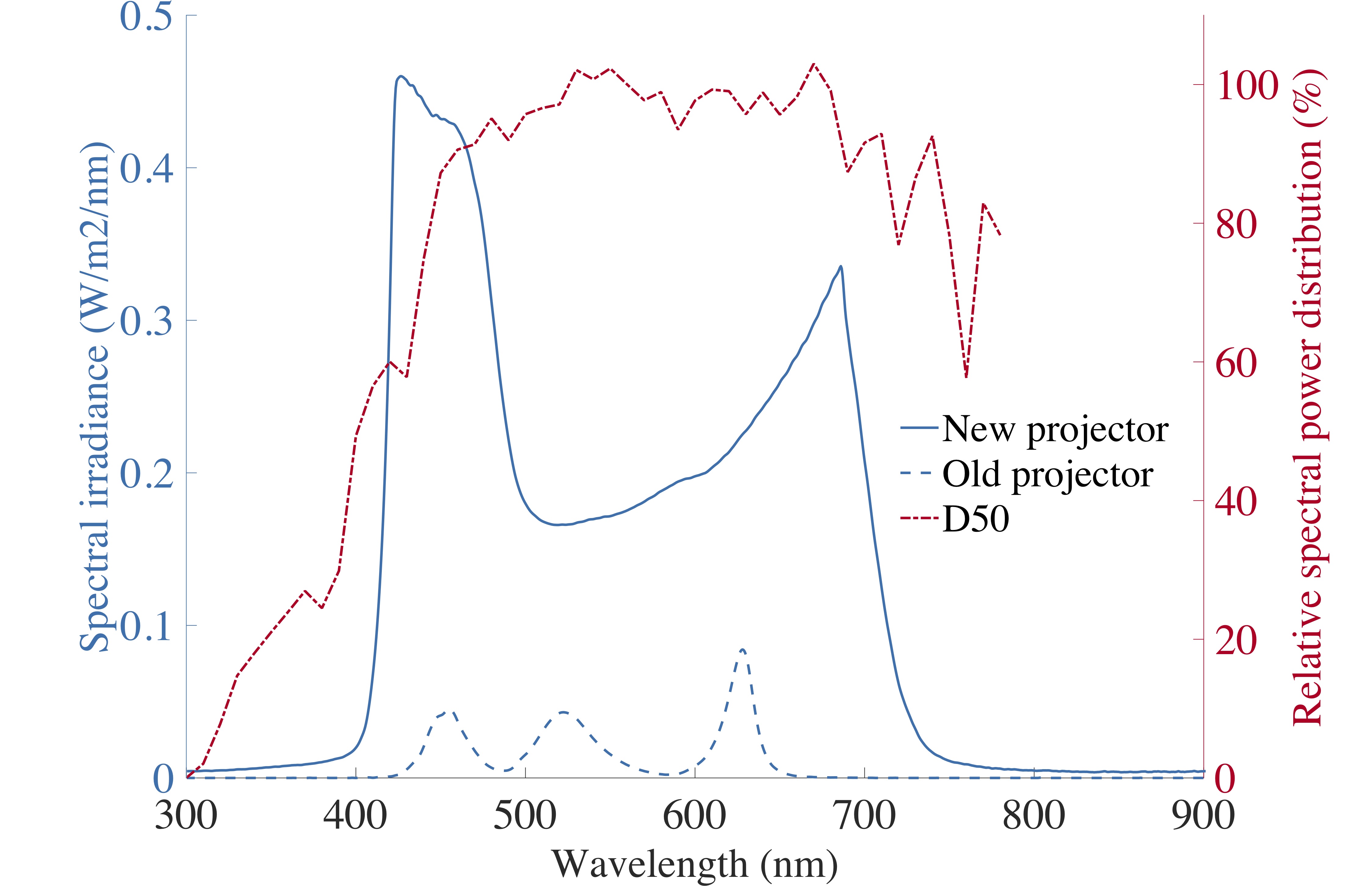}
  \caption{Spectral irradiance of new projector with high pressure mercury lamp illuminant (Acer X133H, solid blue line) in comparison to the old projector with RGB LED illuminant (Optoma PK301 pico-projector, dashed blue line), both with a polarization filter; and relative spectral power distribution of the CIE standard illuminant D50 (solid red line), normalized to a value of 100 at a wavelength of 560 nm \cite{ISO2007}.}
  \label{fig:SpectralPlot}
\end{figure}

\subsubsection{3D scanning module}
The major components for the 3D scanning module are a projector (Acer X113H) and two cameras (Nikon D800E) (see Figure \ref{fig:Scanner}), all these fitted with polarization filters (Hoya HD)to eliminate reflections (cross-polarization). The cameras are fitted with Scheimpflug lenses (Nikkor PC-E 85mm), to align the focal plane of the cameras with the painting surface. The cameras have a resolution of 7424x4924 pixels, and the projector a resolution of 800x600 pixels. Through defocusing the projector slightly, a continuous sinusoidal fringe pattern is projected, thereby not limiting the effective scanning resolution. The cameras capture an area of roughly 180x100mm at the time, hereafter referred to as a \textit{tile}.

The 3D scanning module has nearly the same configuration as described by Zaman et al.\cite{Zaman2014}, but the RGB-LED pico-projector was replaced by a projector using a high-pressure mercury lamp as light source (Acer X113H). Figure \ref{fig:SpectralPlot} shows that the absolute irradiance is much larger for the new projector compared to the old projector (allowing an increased shutter speed, which in turn gives a better signal-to-noise ratio), and that the radiant energy of the new projector covers the full visible spectrum (390-700nm), and thereby better approaches the spectrum of CIE standard illuminant D50 (improving color rendition) \cite{ISO2007}. 
Based on conservation guidelines \cite{Michalski2016} it is estimated that the level of illumination of the projector (9.0 KLux) would lead to a 'Just Noticeable Change' (JCH) on the painting (categorized as medium sensitivity, category 5) after 1 to 3.5 months of exposure. As the exposure is limited to several minutes for any given area, it is estimated to have a minimal impact, similar to the exposure during photography, or during restoration or treatment. The scanner was safely applied, scanning two authentic Dutch, Golden Age paintings, from the collection of the Mauritshuis \cite{Elkhuizen2018a, Mauritshuis2018}.

Although the configuration of components and working principle of our system in terms of 3D topography capture is similar to commercially available structured light scanners, like the Atos Compact Scan \cite{GOM2018}, our scanner has a higher resolution (using 40Mp sensors versus 12Mp camera sensors), meaning that our system can capture a higher resolution for the same capture area. An advantage of using RGB sensors in our setup is the ability to directly register the color information on the 3D data, as they are captured simultaneously. Additionally, the use of Scheimpflug lenses and polarization filters is specifically tuned for the application of scanning planar objects. The alignment of the focal plane to the object plane, means the scan can be made with a larger aperture, leading to faster exposure times. Cross-polarization makes it possible to scan the highly reflective surfaces of paintings. Ultimately, an open system makes it possible to calibrate and register images of the third camera, which is used to capture the spatially varying gloss.

\subsubsection{Gloss scanning module}
The gloss scan module consists of an LED panel (Bresser SH-900, 280x280mm) with a diffuser (4mm translucent Plexiglas), a camera (Nikon D800E, also with a Nikkor PC-E 85mm lens) and stepper motor driving the rotation of a polarization filter, which is mounted on the camera lens. 

\subsubsection{System integration and image processing}
Both scanning modules are mounted to a platform, which in turn is movable horizontally and vertically along the frame, for scanning paintings with maximum dimensions of 1.3x1.3m. All components are controlled via an Arduino\textsuperscript{\textregistered} micro-controller. The scanner is manually positioned in front of the painting. Two distance meters are used to achieve the best possible parallel alignment between the painting and projection plane (measuring four corners of the painting) and for positioning the painting at the right distance, thereby making sure that the whole surface is in focus. 
Camera calibration, image processing and stitching is done using a self-developed software based on Matlab\textsuperscript{\textregistered} 2017a. A Spectralon\textsuperscript{\textregistered} panel (300x300mm) and color calibration target (X-rite Colorchecker\textsuperscript{\textregistered} SG) are used for calibration of the color images. 

\subsection{Printing system} \label{subsec:Materials-Printer}
An adapted version of Oc\'{e} Technologies \cite{OceTechnologiesWebsite2017} \textit{Elevated Printing} technology \cite{ProjectEigerWebsite2017} is used for printing. The ink-jet system utilizes UV-curable inks. A transparent ink is added to the default CMYK and White ink channels, which can be used to create spatially-varying gloss. This experimental printer has a build volume of \SI{1.25x2.5}{\meter} (X,Y) and \SI{5}{\milli\meter} (Z) height. The printing system has a planar resolution of 450dpi (\SI{56}{\micro\meter}), and the smallest possible layer resolution is \SI{2}{\micro\meter}. 

\section{Method} \label{sec:Method}

\subsection{Calibration and scanning workflow}
\begin{figure*}
\centering
\includegraphics[width=1\textwidth]{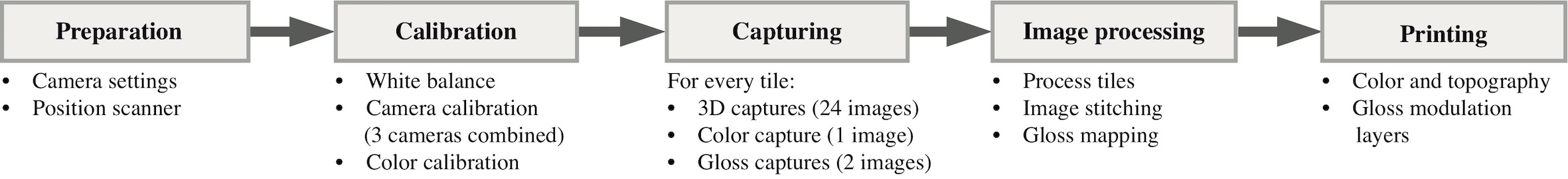}
\caption{Overview of workflow for the reproduction of color, gloss and topography.}\label{fig:workflow}
\end{figure*}

Figure \ref{fig:workflow} presents the workflow for calibration and scanning of a painting when the scanner is assembled. Firstly, all devices are set to the appropriate settings and the white balance, the color and the lens distortion of three cameras are calibrated based on multi-view geometry \cite{Bouguet2013}. Images are then captured for 3D, color and gloss reconstruction regarding each tile. After scanning the color, topography and gloss images are processed off-line for each tile, and then they are stitched to form the color, topography and gloss images of the whole painting. The gloss image is mapped to the printable gloss range, and finally a 3D print is made using the color, topography and gloss map. Details on color and topography capture, gloss capturing and fabrication are presented in the following sections.

\subsection{Color and topography capturing} \label{subsec:Method-ColTopCapt}
The color and 3D topography of the surface are captured using a hybrid solution of fringe projection and stereo imaging (described in \cite{Zaman2014}). 
A 6-phase shifting sinusoidal greyscale pattern (fringe) is projected horizontally and vertically to acquire 3D information of the projected area (24 images). Fringes are unwrapped and a sparse stereo matching is employed to match the fringes of both camera images. Once the fringes are matched a lookup table is generated for both cameras, encoding both images. Finally a dense stereo matching is made, using the ray-tracing principle, taking into account the camera calibration.
One additional image is captured with a uniform illumination of the projector, which is used as the color image. With the frontal illumination of the projector, shadowing is very limited and free of specular reflections (due to the cross-polarization). Also the effect of local shading (whereby the captured color is actually the product of the diffuse albedo and the dot product between the surface normal and the normalized lighting direction) is minimal, not leading to visual artifacts. In short the color image can be used as a proxy for the diffuse albedo. The color image is corrected for the non-uniform illumination of the projector (i.e. caused by lens vignetting), using the frame-filling calibration image captured of the Spectralon\textsuperscript{\textregistered} panel.
The current setup is able to achieve an in-plane resolution of \SI{25x25}{\micro\meter} (XY), outputting an RGB color image and a height map. 

\subsection{Gloss capturing} \label{subsec:Method-GlossCapt}
\begin{figure}
\centering
  \includegraphics[width=0.5\textwidth]{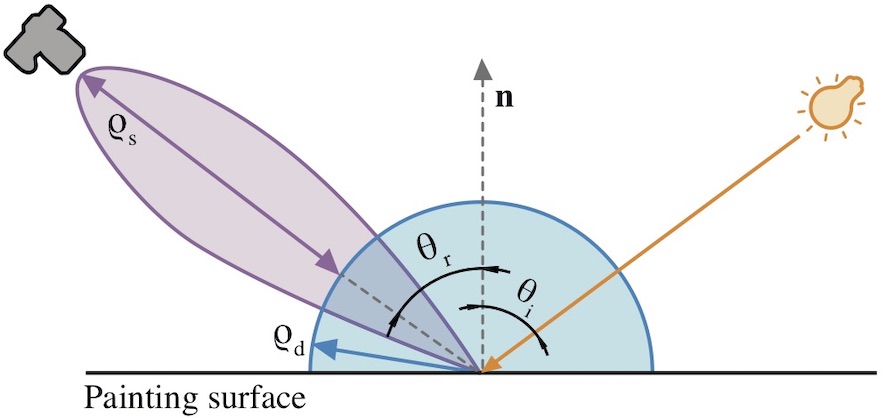}
  \caption{Dichromatic reflection model, where $\theta_{i}$ is the incidence illumination angle and $\theta_{r}$ the reflection angle, $\rho_{d}$ and $\rho_{s}$ represent the diffuse and specular reflectance components.}
  \label{fig:DichromaticReflModel}
\end{figure}

Following Shafer's Dichromatic Reflection Model \cite{Shafer1984}, we assume that the surface reflectance can be modeled with a diffuse 
(or 'body') and specular (or 'interface') reflectance component, as a function of the incident angle ($\theta_i$), reflectance angle ($\theta_r$), the phase angle ($\phi$), and the wavelength ($\lambda$):

\begin{equation}\label{Eq:TotalMeasuredIntensity}
\begin{split}
L(\theta_i, \theta_r, \phi, \lambda) & = L_{s} (\theta_i, \theta_r, \phi, \lambda) + L_{d} (\theta_i, \theta_r, \phi, \lambda) \\
& = \rho_{s} (\theta_i, \theta_r, \phi) c_{s}(\lambda) + \rho_{d} (\theta_i, \theta_r, \phi)c_{d}(\lambda)
\end{split}
\end{equation}

\noindent where $\rho_{d}$ represents the magnitude of the diffuse component and $\rho_{s}$ of the specular component, which depends only on geometry and is independent of wavelength. 
Figure \ref{fig:DichromaticReflModel} shows a schematic representation of Shafer's BRDF model, showing the incident illumination angle $\theta_{i}$, the reflection angle $\theta_{r}$ (at the specular peak), and the reflectance components $\rho_{d}$ and $\rho_{s}$.
The magnitude of the specular component ($\rho_{s}$) is extracted across the surface, varying in relative intensity due to micro-scale roughness scattering (low intensity for a rough surface and high for a glossy surface), as the input parameter for the gloss printing.

\subsubsection{Brewster's angle reflectance polarization}
In the proposed approach the reflectance is sampled at the mirror reflection angle ($\theta_\mathrm{i} = \theta_\mathrm{r}$). We assume that the magnitude of the specular reflectance is influenced by the scattering effect due to the surface roughness (the refractive index is similar across the surface, and the illuminance and Fresnel effects are normalized). 
In perception experiments it was found the roughness parameter is a good predictor of the perceived glossiness \cite{Ferwerda2001,Fleming2003}. Here we assume that the specular peak reflectance of a point is dominant over the contribution of the spread of the specular lobe of neighboring points.  

The proposed approach also assumes that most of the painted surface is sufficiently flat that the measurement is close enough to the specular peak (or locally corrected for using the height map, see section \ref{subsubsec:SurfNormShadow}) for reproducing the spatially-varying gloss characteristics of a painting's surface. 

Extracting the specular reflectance component is achieved by utilizing the polarization of reflections. The intensity and polarization of reflections, can be calculated using the Fresnel equations \cite{Hecht2002}:
\begin{equation} \label{Eq:FresnelRsRp}
R_\mathrm{s}(\theta) = \left(\frac{n_1\cos\theta_\mathrm{i}-n_2\cos\theta_\mathrm{t}}{n_1\cos\theta_\mathrm{i}+n_2\cos\theta_\mathrm{t}}\right)^2 
\qquad \text{and} \qquad
R_\mathrm{p}(\theta) = \left(\frac{n_1\cos\theta_\mathrm{t}-n_2\cos\theta_\mathrm{i}}		{n_1\cos\theta_\mathrm{t}+n_2\cos\theta_\mathrm{i}}\right)^2
\end{equation}

\noindent where reflection coefficients $R_\mathrm{s}$ and $R_\mathrm{p}$ correspond to the perpendicular (Senkrecht, in German) and parallel directions to the surface. $n_1$ and $n_2$ are the refractive indexes of air and the material being scanned. $\theta_\mathrm{i}$ and $\theta_\mathrm{t}$ are the incident and transmission angles, whereby the latter can be substituted using Snell's law and trigonometric identities \cite{Hecht2002}: 
\begin{equation}\label{Eq:SnellsLaw}
\frac{n_2}{n_1} = \frac{sin(\theta_\mathrm{i})}{sin(\theta_\mathrm{t})} \quad \Rightarrow \quad  \cos(\theta_\mathrm{t})=\sqrt[]{1-\left(\frac{n_1}{n_2}\sin(\theta_\mathrm{i})\right)^2}
\end{equation}

Hereof, the incident angle where the reflection is fully polarized ($R_\mathrm{p} = 0$), called Brewster's angle, can be calculated as follows \cite{Hecht2002}:
\begin{equation}\label{Eq:BrewsterAngle}
\theta_\mathrm{B} = \arctan(\frac{n_2}{n_1})
\end{equation}

\begin{figure}
\centering
\begin{minipage}[t]{.55\textwidth}
\begin{center}
\vspace{0pt} 
  \includegraphics[width=\linewidth]{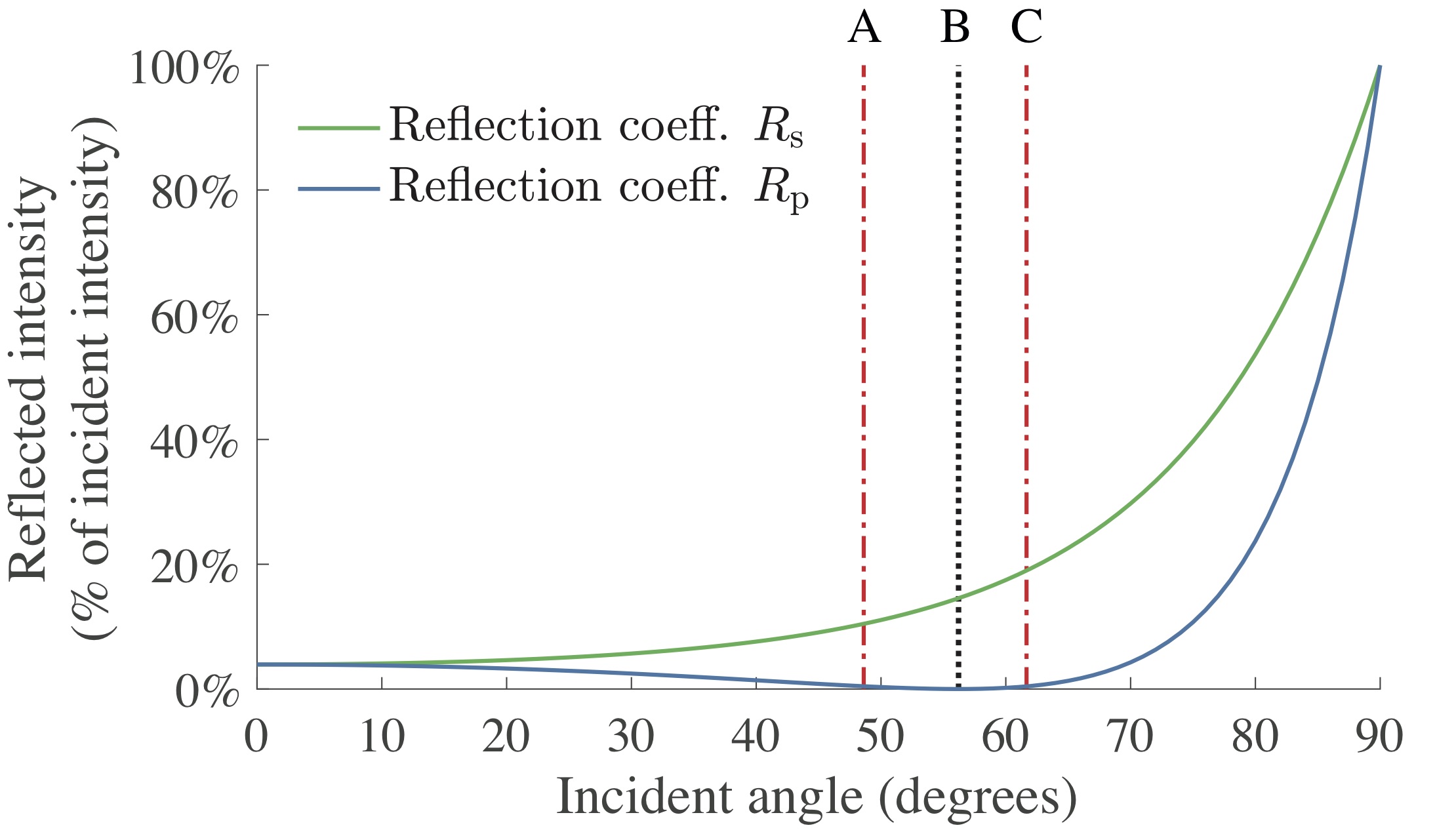}
  \captionof{figure}{Reflection coefficients $R_\mathrm{s}$ and $R_\mathrm{p}$ are plotted for the average refractive index ($n_2$=1.495) for every incident angle. The plots show the point of maximum polarization at Brewster's angle (black dotted line), and the mirror reflection angles at the edge of the scan region at point A and C (red dashed lines, see figure \ref{fig:ScannerDiagram})}
  \label{fig:BrewsterAnglePlot}
\end{center}
\end{minipage}%
\hfill
\begin{minipage}[t]{.42\textwidth}
\begin{center}
\vspace{0pt} 
  \includegraphics[width=\linewidth]{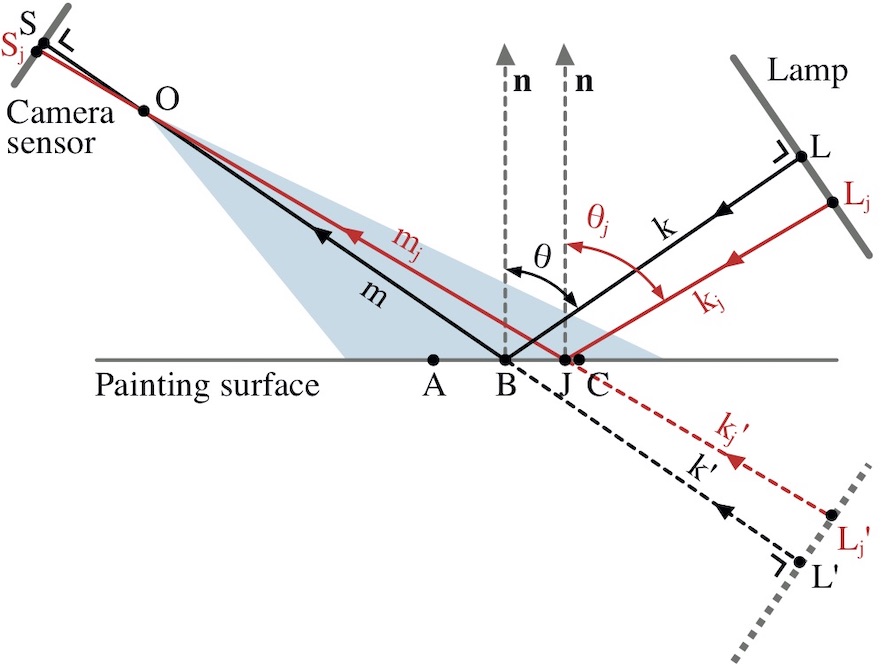}
  \captionof{figure}{Schematic top view of gloss scanning module, consisting of a lamp (right) and camera (left) both rotated by Brewster's angle ($\theta$) w.r.t. the surface normal ($\vec{n}$). Point B denotes the scan area center, and point $A$ and $C$ the 3D scan area boundaries, and blue shaded region represents the gloss camera field-of-view. Any point ($J$) on the surface has a mirror reflection angle ($\theta_j$) with distances $k_{j}$ to the lamp and $m_{j}$ to the camera optical center $O$ ($k_{j}'$ is the distance from the surface to the Lamp's mirror image).}
  \label{fig:ScannerDiagram}
\end{center}
\end{minipage}
\end{figure}

Oil paints and varnishes that are typically used in oil paintings have a refractive index in the range of 1.47 to 1.52 \cite{DelaRie2010}, yielding a Brewster's angle between \SI{55.8}{\degree} and \SI{56.7}{\degree}. The averaged refractive index ($n_2=1.495$), which gives a Brewster's angle ($\theta$) of \SI{56.3}{\degree}, is adopted in the setup of the system and data processing (for estimations of errors due to refractive index variation see table \ref{tab:RsRpRatios}). The relationships between reflection coefficients ($R_\mathrm{s}$ and $R_\mathrm{p}$) and the incident angle are plotted in figure \ref{fig:BrewsterAnglePlot} (for $n_2=1.495$). As reflection coefficient $R_\mathrm{p}$ goes to zero at Brewster's angle, the light is effectively polarized at (and for the most part around) this angle. This effect ($R_p\approx0$) can be used to discriminate the specular reflectance from the diffuse reflectance. 

\subsubsection{Area capture of specular reflectance using polarization}
In the setup of the scan modules (see figure \ref{fig:ScannerDiagram}), a diffuse light source which is a flat panel illuminates the surface of the painting ($k$=\SI{450}{\milli\meter}). The camera captures the illuminated area. Given point $B$ in the painting, which is the center of the scan area, the angle $\theta$ between the plane normal ($\vec{n}$) and the perpendicular direction of the light source ($k$) is set as Brewster's angle ($\theta=\SI{56.3}{\degree}$). The angle between the plane normal ($\vec{n}$) and the optical axis of the camera ($m$) is set as Brewster's angle as well. The width of a single gloss capture is approximately \SI{180}{\milli\meter} on the painting (distance between point $A$ and $C$), corresponding to the field-of-view of the 3D scan module. The distances are estimated based on the camera calibration and measurements using a laser distance meter.

Given our assumption that the specular peak reflection is dominant over the spread of the specular lobe, we simplify the reflection model that any point ($J$) has a corresponding mirror reflection angle ($\theta_j$), at distance ($k_j$) to the lamp and distance ($m_j$) to the camera. Note that the light source is sufficiently large to include the mirror angle for every surface point within a capture region. In figure \ref{fig:BrewsterAnglePlot} the region between the red-dashed lines indicates the mirror angle range (and their corresponding coefficient ranges) that are present in a single capture. To extract the specular reflectance, two images are captured, one image containing specular reflections ($I_1$) and one image where the specular reflections are filtered by rotating the polarization filter in front of the camera lens by \SI{90}{\degree} ($I_2$). The images are converted from RGB to HSL color space, where the lightness channel (L) is used for further image processing, safely neglecting the wavelength dependency of the specular reflectance. Here we assume the paint behaves as a dielectric material, and therefore does not alter the spectral properties of the reflections \cite{Klinker1987}. The difference between these two images is calculated:
\begin{equation}\label{Eq:IntensityDifference}
I_\mathrm{g} = I_1(\rho_\mathrm{d} + \rho_\mathrm{s}) - I_2(\rho_\mathrm{d} + \rho_\mathrm{s}) 
\end{equation}
\noindent As the diffuse reflectance is unpolarized, the contribution of the diffuse reflectance will be identical in both images ($I_1(\rho_\mathrm{d})$ = $I_2(\rho_\mathrm{d})$) and from the Fresnel equations follows that specular component in the second image goes to zero ($I_2(\rho_\mathrm{s}) \approx 0$). Therefore $I_\mathrm{g}$ is the gloss map ($I_\mathrm{g}$ = $I_1(\rho_\mathrm{s})$), giving the relative magnitude in specular gloss across the surface.

\subsubsection{Gloss calibration} \label{subsubsec:GlossCalib}
If we consider the surface to act like a mirror (either highly reflecting or scattering) due to the angular configuration, the distance between the lamp and camera is not constant over the surface (i.e. ($k' + m) \neq (k_j' + m_j$), see figure \ref{fig:ScannerDiagram}). Additionally, as the mirror reflection angle varies across the surface, the reflected intensity also varies, following from the Fresnel equations (see $R_\mathrm{s}$ in figure \ref{fig:BrewsterAnglePlot}). The gloss map is rescaled relative to the center of the image, taking into account both effects. 

Although it seems easy to correct the irradiance variation using the image of a Spectralon\textsuperscript{\textregistered} panel illuminated by the LED panel as a reference object, such a image cannot capture the correct irradiance variation. It would give the illumination sum from the whole LED panel for any surface point, rather than just the illumination arriving from the mirror reflection angle. For this reason a model based correction better represents the irradiance variation, applicable for the specular reflectance component.

First, the measured gloss map ($I_\mathrm{g}$) is scaled relative to the center of the image (at $B$) to correct for the variation in irradiance arriving at the sensor from the lamp, reflected by any point ($J$), applying the inverse-square law \cite{Hecht2002}:

\begin{equation}\label{Eq:IllumCor}
\frac{E(S\mathrm{j})}{E(S)}=\frac{({k+m})^2}{({k_\mathrm{j}+m_\mathrm{j}})^2} \quad \textrm{thus,} \quad E(S\mathrm{j}) = E(S) \frac{({k+m})^2}{({k_\mathrm{j}+m_\mathrm{j}})^2}
\end{equation} 
\noindent where $E(S)$ and $E(S\mathrm{j})$ are the irradiance at points $S$ and $S_\mathrm{j}$ on the sensor, corresponding to point $L$ and $L_\mathrm{j}$ on the lamp, reflected at point $B$ and $J$, respectively. 

Secondly, the intensity is scaled for the radiance specularly reflected at every point, relative to the center of the scan, by using the difference between reflection coefficients as scaling factors:
\begin{equation}\label{Eq:ReflCoeffCor}
E(S_\mathrm{j})= E(S) \frac{R_\mathrm{s}(\theta)-R_\mathrm{p}(\theta)}{R_\mathrm{s}(\theta_j)-R_\mathrm{p}(\theta_j)}
\end{equation}
\noindent where $E(S)$ and $E(S_\mathrm{B})$ is the irradiance measured at point $S$ and $S_\mathrm{j}$ on the sensor, corresponding to the reflection at point $B$ and $J$, respectively. $R_\mathrm{s}$ and $R_\mathrm{p}$ are reflection coefficients as a function of the incident angle (see equation \eqref{Eq:FresnelRsRp} and figure \ref*{fig:BrewsterAnglePlot}). This describes the relationship in reflected intensity, assuming equal irradiance and equal scattering.

As the measured irradiance is proportional to the intensity as well as the reflection coefficient ratio, the above formulas can be translated to scaling factors:
\begin{equation}
I_\mathrm{g,cor}(J) = e * f * 
I_\mathrm{g}(J)
\quad \textrm{where} \quad
e = \frac{({k+m})^2}{({k_\mathrm{j}+m_\mathrm{j}})^2}
\textrm{, } \quad
f = \frac{R_\mathrm{s}(\theta)-R_\mathrm{p}(\theta)}
{R_\mathrm{s}(\theta_j)-R_\mathrm{p}(\theta_j)} 
\end{equation}

\noindent where $I_\mathrm{g,cor}(J)$ in the corrected gloss intensity of point $J$, and $e$ and $f$ are scaling factors relating to the varying irradiance and reflection coefficient ratios. The output of the gloss scanning is a gray-scale gloss map, scaled between the minimum and maximum measured gloss intensity. 
Based on the camera calibration matrix, using multi-view geometry\cite{Bouguet2013}, the gloss image is mapped to the height map. 

Table \ref{tab:RsRpRatios} shows the reflection coefficients $R_\mathrm{s}$ and $R_\mathrm{p}$ at the mirror reflection angles of points $A$, $B$ and $C$, for the minimum, average and maximum refractive index, found in oil paintings. For the average refractive index ($n_2$=1.495), 4.1\% and 2.2\% of the reflected light remains unpolarized at the boundaries of the image ($I_2(\rho_\mathrm{s}) \neq 0$ in equation \eqref{Eq:IntensityDifference}). The theoretical maximum measurement error is 4.5\% (found at point $A$), when a painted surface with a higher refractive index ($n=1.52$) is imaged at the chosen configuration. In the calculation of the scaling factors in equation \eqref{Eq:ReflCoeffCor}, the error introduced by variation of refractive index is less than 1\%. Both errors are neglected in our approach.

\begin{table}
\begin{center}
\caption{Reflection coefficients $R_s$ and $R_p$ (see Equation \eqref{Eq:FresnelRsRp}), and percentage of reflection that remains unpolarized, for minimum, average and maximum refractive index found in oil paintings, when the scanner is configured at the average Brewster's angle ($\theta$=\SI{56.3}{\degree}), at the left image boundary ($A$), center ($B$) and right image boundary ($C$).} \label{tab:RsRpRatios}
\begin{tabular}{llllllllll} 
  \cmidrule{2-10}
  &  \multicolumn{3}{c}{$n_2$=1.47} & \multicolumn{3}{c}{$n_2$=1.495 (average)}   & \multicolumn{3}{c}{$n_2$=1.52}\\
  \cmidrule(r){2-4}\cmidrule(r){5-7}\cmidrule(r){8-10}
  & $A$ & $B$ & $C$ & $A$ & $B$ & $C$ & $A$ & $B$ & $C$\\
\midrule
$R_\mathrm{p}$ coefficient &0.0037 &\num{2.08e-5} &0.0047 &0.0044 &0 	 &0.0042 &0.0052 &\num{2.18e-5} &0.0037 \\
$R_\mathrm{s}$ coefficient &0.098  &0.138 		 &0.181	 &0.1045 &0.1456 &0.1899 &0.111  &0.153 		&0.199 \\
$R_\mathrm{p}$/($R_\mathrm{p}$+$R_\mathrm{s}$) error (\%)	&3.66 &0.015 &2.18 &4.06 &0 &2.18 &4.47 &0.014 &1.84\\
\bottomrule 
\end{tabular}
\end{center}
\end{table}

\subsubsection{Surface normal and shadow correction} \label{subsubsec:SurfNormShadow}
\begin{figure}
\centering
\begin{minipage}{.5\textwidth}
  \centering
\includegraphics[width=0.9\linewidth]{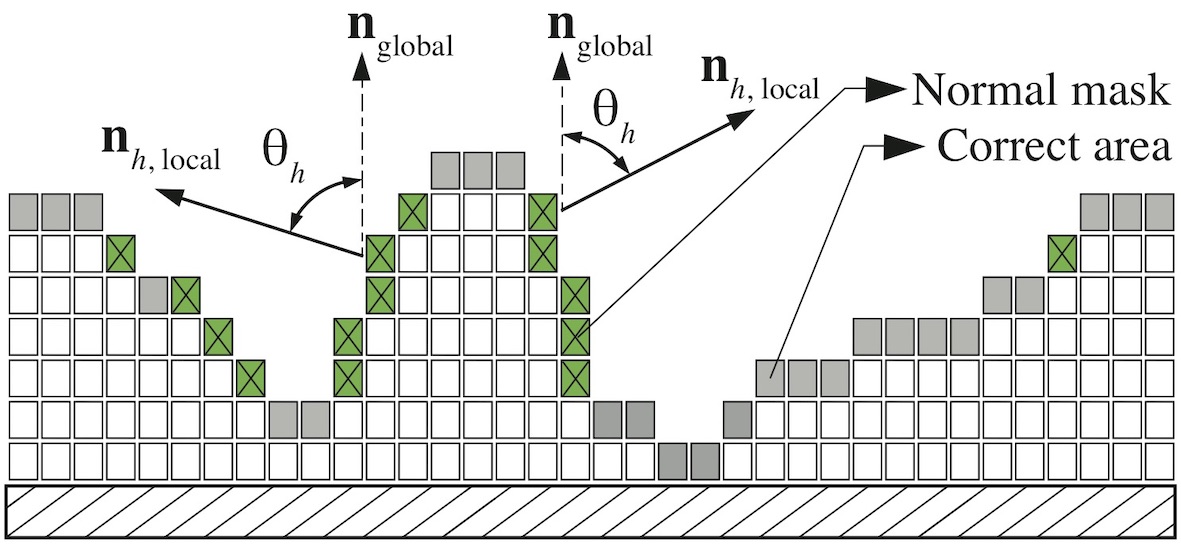}
\subcaption{Surface normal mask} \label{subfig:NormalMask}
\end{minipage}%
\begin{minipage}{.5\textwidth}
  \centering
\includegraphics[width=0.9\linewidth]{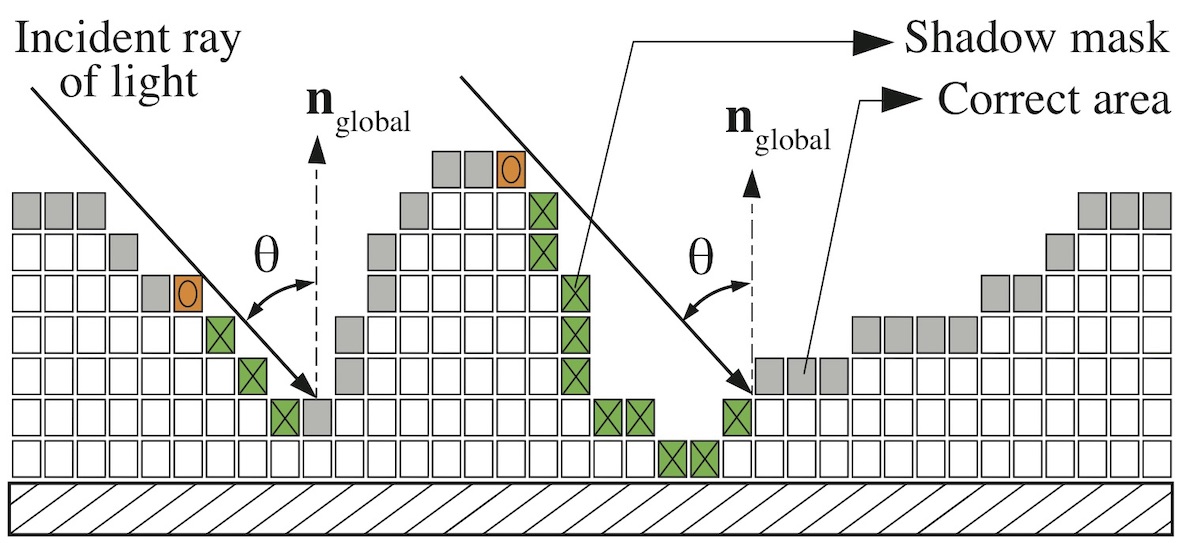}
\subcaption{Shadow mask} \label{subfig:ShadowMask}
\end{minipage}%
\caption{(a) Surface normal mask, showing the angle ($\theta_h$) between the local ($\vec{n}_{h, \mathrm{local}}$) and global surface normal ($\vec{n}_\mathrm{global}$) and the pixels (green, crossed) belonging to the mask, which exceed an experimentally determined threshold; (b) Shadow mask, showing Brewster's angle ($\theta$) between incident ray of light and the global surface normal ($\vec{n}$), the critical edge pixel (orange, oval) and the pixels belonging to the mask (green, crossed).} \label{fig:Masks}
\end{figure}

Due to the 3D structure of the painting, the surface normal varies locally, and therefore some parts of the painting are not captured at the (Brewster's) mirror angle. Utilizing the height map, the local surface normal is determined, and regions which exceed an experimentally determined threshold are masked, as illustrated by the green, crossed pixels in figure \ref{subfig:NormalMask}. The normal mask threshold was experimentally determined at $\theta_h>\SI{10}{\degree}$, where we saw a sharp transition in gloss intensity, in regions where we expect a continuous glossiness. Additionally, the incident angle of light combined with 3D structure, causes shadows and shaded regions. An additional shadow mask is created, also utilizing the height map (illustrated by the green, crossed pixels in figure \ref{subfig:ShadowMask}). The surface normal and shadow mask are combined and in the data processing, gloss information in the masked regions is discarded and filled using the local maximum value. 
For every pixel, the local maximum value is determined, sampling the closest pixels (in a radius of 40 pixels). These values are used to fill the pixels which fall within the mask. Here we assume some local continuity in gloss, and that the local maximum is produced by the region which is measured at the mirror angle (the local surface normal coincides with the global normal). Alternatively, more sophisticated interpolation algorithms (for instance a Poisson infilling strategy \cite{Perez2003}) might also be applied here, to minimize visual artifacts.

\subsection{Tile stitching}
Images are stitched using a self-developed algorithm. First, based on the assumption that a painting is generally a planar surface, a best-fit plane is found for the center tile. Then the 3D topography data of each tile is fitted to this plane individually and the transformation matrix of each tile is documented. As the 3D topography, color and gloss are aligned for each tile, color images and gloss images of this tile can be transformed to fit this plane by the same transformation matrix. Then, adjacent tiles are matched using a cost function which features the RGB color as well as 3D information. The images are first matched and stitched in rows, after which the rows are stitched to form the complete image. 
The height data is aligned using the mean height of the overlapping regions, and a (residual) slope difference is removed. The data is then merged, blending the low frequency variations, whilst superimposing the high frequency variations of one of both images, leading to satisfactory stitching with minimal visual artifacts. More sophisticated merging, for instance using minimal deformation strategies, might further improve stitching.

\subsection{Appearance fabrication}\label{subsec:Method-Fabrication}
\begin{figure}
\centering
\begin{minipage}[t]{.6\textwidth}
\begin{center}
  \includegraphics[width=0.9\linewidth]
  {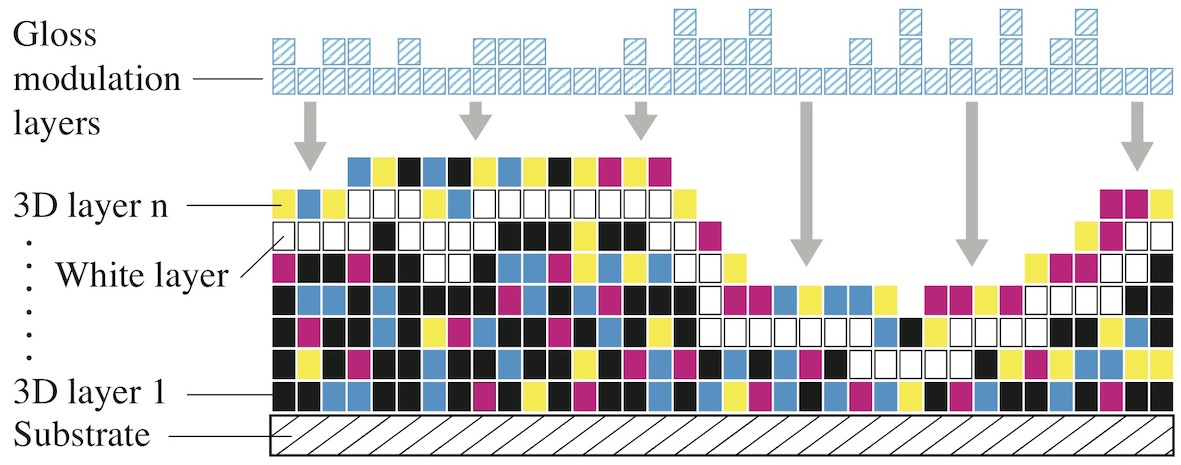}
\captionof{figure}{Cross section diagram of 3D print showing the stratigraphy of the substrate, topography/color layers with intermediate white layer, and gloss layers}
  \label{fig:AppearanceFabrication}
  \end{center}
\end{minipage}%
\hfill
\begin{minipage}[t]{.35\textwidth}
  \begin{center}
  \includegraphics[width=\linewidth]
  {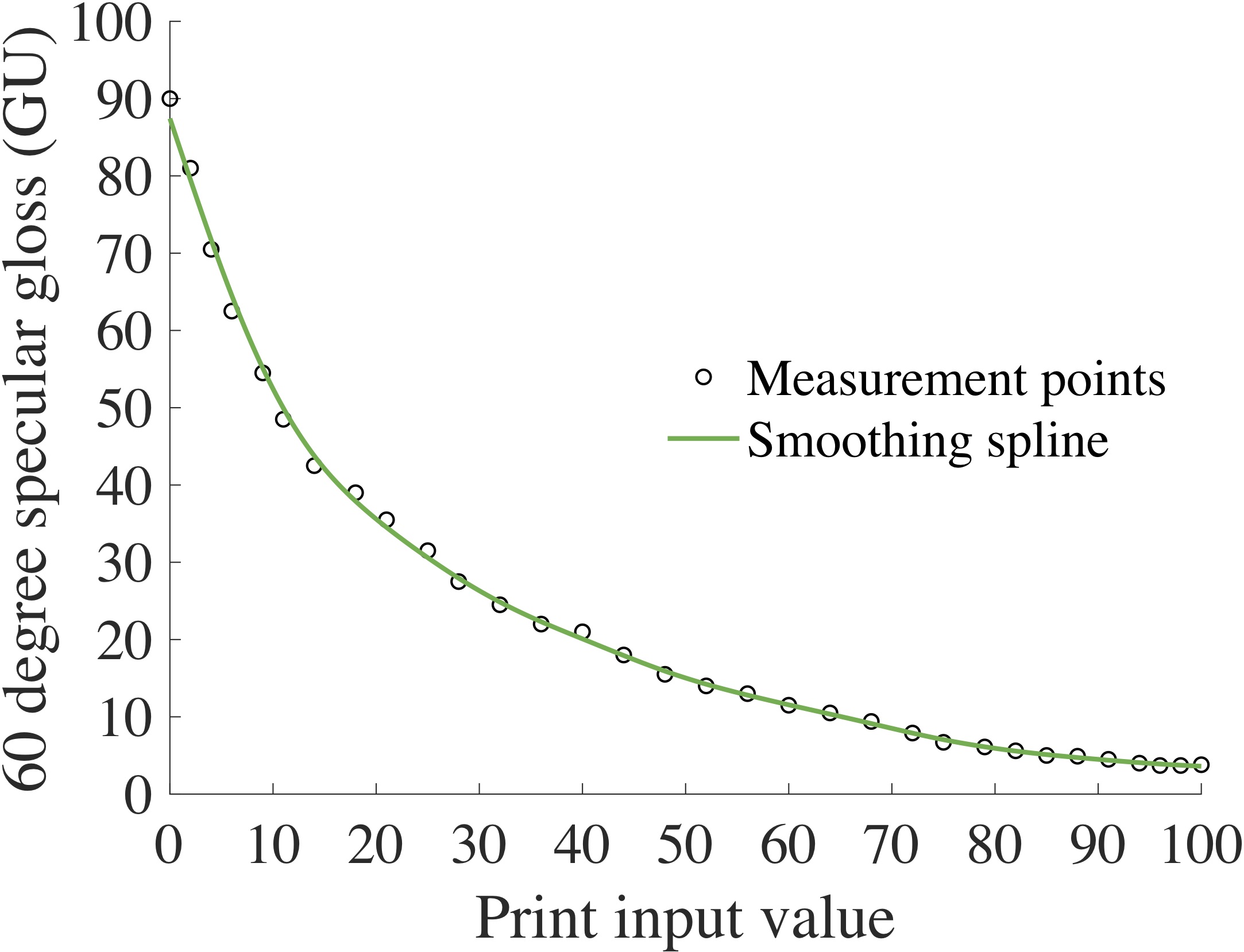}
  \captionof{figure}{Relationship between print input values and \SI{60}{\degree} specular gloss in gloss units (GU), measured using a glossmeter.}
  \label{fig:GlossmeterRelationships}
\end{center}
\end{minipage}%
\end{figure}

To fabricate the appearance, three input files are used: a color image, a height map (resembling the 3D topography) and the gloss map. Through the use of Oc\'{e} Technologies custom 3D slicer software, the color image and height map are combined to form a stack of bitmaps. The color of every position (X-Y) is printed in the layer corresponding to the height specified for that position (Z). Consequently, the voxels below that point are set to print in white ink, to form a boundary between the bulk of the print (printed with all color channels) and the outer color layer, as the color printing is based on subtractive color mixing on a white substrate (which is this case is the white ink). Thus, the color and topography are printed integrally using cyan, magenta, yellow, black and white ink (see figure \ref{fig:AppearanceFabrication}). 

Spatially-varying gloss is created with six consecutive layers of transparent ink, printed on top of the 3D color layers (see figure \ref{fig:AppearanceFabrication}). The first layer is a high gloss layer; the ink is applied in a full coverage and left to flow across the surface before it is cured, hereby creating the highest printable gloss level. This high gloss layer across the whole surface ensures the creation of a smooth gloss gradient, from high gloss to matte. In the following layers an input gloss gradient is dithered, so that the matte part receives the most ink coverage. 
Each of these 6 layers are cured directly after printing, creating a rougher surface finish. Sample gloss patches were printed (given a printer input value between 0-100), and measured with a glossmeter (Byk Micro-tri-glossmeter) to determine their glossiness. The relationship between the printer input values (0-100\% gloss) and \SI{60}{\degree} specular gloss, is depicted in figure \ref{fig:GlossmeterRelationships}. As this relationship is non-linear, the gloss map is multiplied by the inverse of the fitted function, as to create a linear gloss mapping to the glossmeter values: 
\begin{equation}
P = f(G_{60})^{-1} * I_\mathrm{g, cor}
\end{equation}
\noindent where, $P$ is the print value, $f(G_{60})$ is the function that described the relationship between print values and \SI{60}{\degree} specular ($G_{60}$) gloss, and $I_\mathrm{g, cor}$ is the corrected glossmap. The gloss map, scaled between the minimum and maximum measured value, is mapped to the full printable gloss scale.

\section{Results} \label{sec:Results}

\begin{figure}
\begin{center}
     \setlength{\tabcolsep}{0pt}
\newcolumntype{C}{>{\centering\arraybackslash}p{0.16\linewidth}}
     \begin{tabular}{CCCCCC} 
\multicolumn{6}{c}{\includegraphics[width=0.95\linewidth]{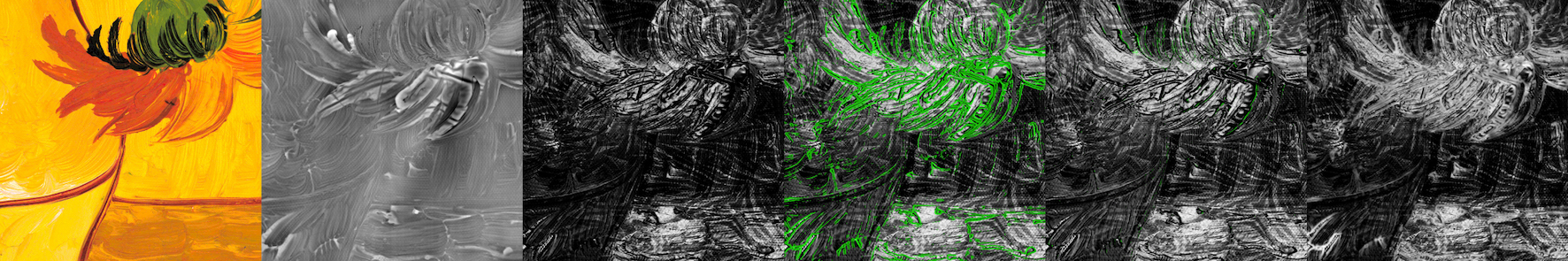}}\\
\multicolumn{6}{c}{\includegraphics[width=0.95\linewidth]{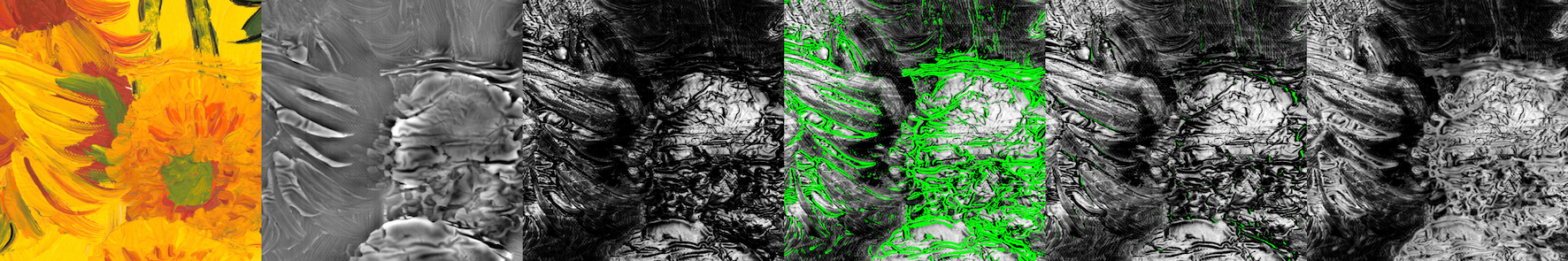}}\\
\multicolumn{6}{c}{\includegraphics[width=0.95\linewidth]{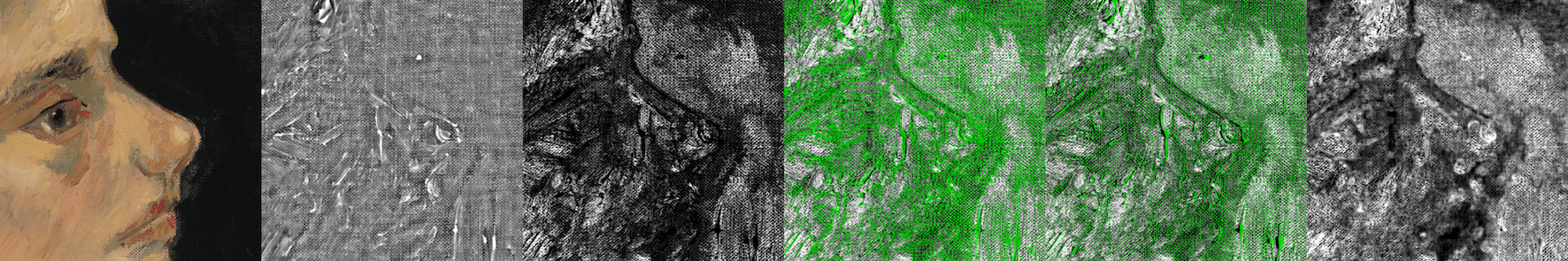}}\\
\multicolumn{6}{c}{\includegraphics[width=0.95\linewidth]{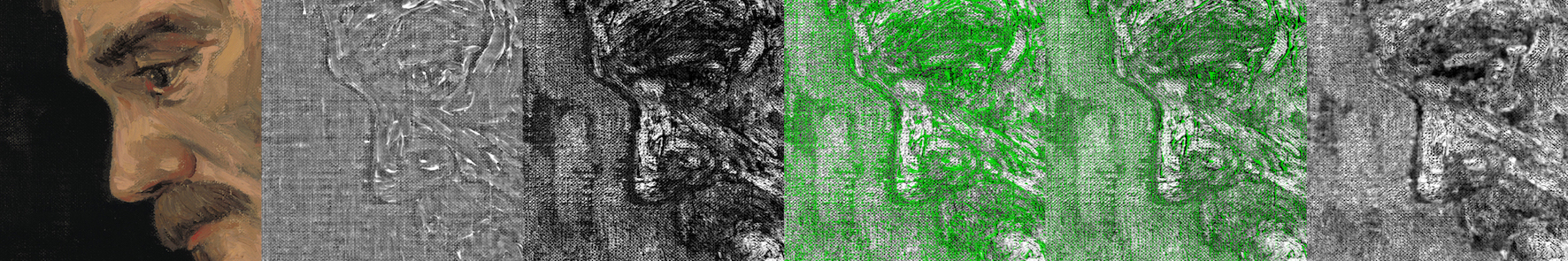}}\\
     {\sf\small (a) Color image}&
     {\sf\small (b) Heightmap}& 
     {\sf\small (c) Glossmap}& 
     {\sf\small (d) Normal mask}& 
     {\sf\small (e) Shadow mask}& 
     {\sf\small (f) Glossmap}\\ 
     {\sf\small }&
     {\sf\small }&
     {\sf\small uncorrected}&
     {\sf\small on glossmap}&
     {\sf\small on glossmap}&
     {\sf\small corrected}\\
\end{tabular}
\end{center}
\caption{Scan results, details of \textit{Sunflowers} (top two rows) and \textit{Two Wrestling Figures} (bottom two rows)} \label{fig:ResultScan}
\end{figure}

\subsection{Scan results} \label{subsec:Scan results}
Two paintings, \textit{Two Wrestling Figures} and \textit{Sunflowers}, both in the style of Vincent van Gogh, were used as a case study for the proposed reproduction workflow. \textit{Two Wrestling Figures} (\SI{99x79}{\centi\meter}) was captured in 140 tiles and the \textit{Sunflowers} (\SI{30x40}{\centi\meter}) in 42 tiles, both with an overlap of approximately 30\%  between tiles to enable stitching. Per tile, 24 images are captured for the 3D reconstruction (12 by each 3D camera), 2 images for color (1 by each 3D camera, also used for the sparse stereo matching), and 2 consecutive images for the gloss reconstruction by the gloss camera. In total 3780 images of 40Mpixel were captured for one scan of the \textit{Two Wrestling Figures}, and 1134 for the \textit{Sunflowers}, of which 7.4\% (280 and 84 images, respectively) are needed for the gloss reconstruction.

With our current implementation, capturing a single tile (including movement to the next position) takes on average 7m50s, meaning that the \textit{Sunflowers} can be captured in 4h18min and the \textit{Two Wrestling Figures} in 18h16m. The images are processed off-line, where reconstructing the 3D image and gloss combined takes approximately 12 minutes per tile. Finally, the tiles are stitched to create the complete image, taking approximately four hours in the case of the \textit{Sunflowers} painting.

Details of the scans are shown in figure \ref{fig:ResultScan}. Figure \ref{fig:ResultScan}(a) shows the RGB color images. The heightmaps, with a maximum height variation of \SI{1.1}{\milli\meter} are shown in figure \ref{fig:ResultScan}(b). The uncorrected glossmap, the shadow and normal mask mask on the glossmap (depicted in green), and the corrected gloss map are shown in figure \ref{fig:ResultScan}(c-f)). For the \textit{Sunflowers} painting, the normal mask covers 13.3\% of the total surface and the shadow mask 0.4\%, where 0.35\% of the surface is masked by both. For the \textit{Two Wrestling Figures}, these percentages are 30.1\% and 22.6\%, where 9.6\% of the surface is masked in both. We found that even for these paintings (with relatively pronounced topography), there is still enough unmasked area to fill in the masked regions. 

\begin{figure}[htb]
    \centering
        \begin{subfigure}[t]{0.23\textwidth}
            \includegraphics[height=1\textwidth]{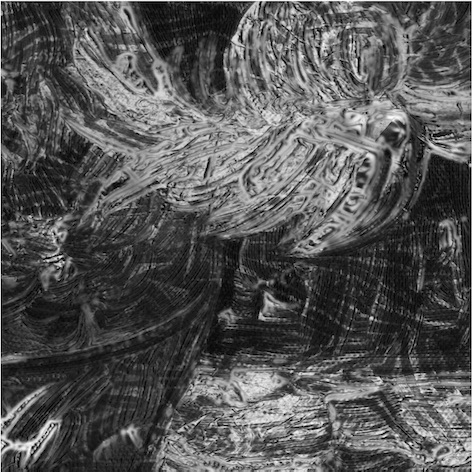}
            \caption{$I_{0}$}
            \end{subfigure}%
        \begin{subfigure}[t]{0.23\textwidth}
            \includegraphics[height=1\textwidth]{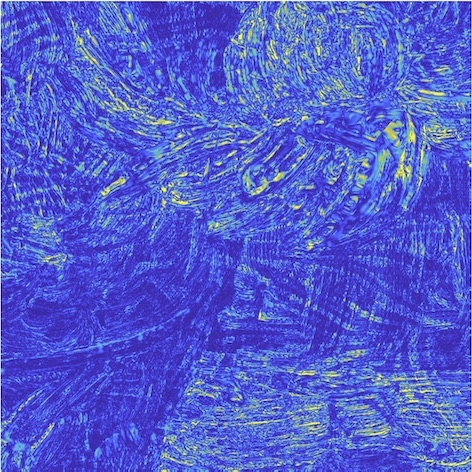}
            \caption{$\lvert I_{90} - I_{0} \rvert$}
        \end{subfigure}%
        \begin{subfigure}[t]{0.23\textwidth}
            \includegraphics[height=1\textwidth]{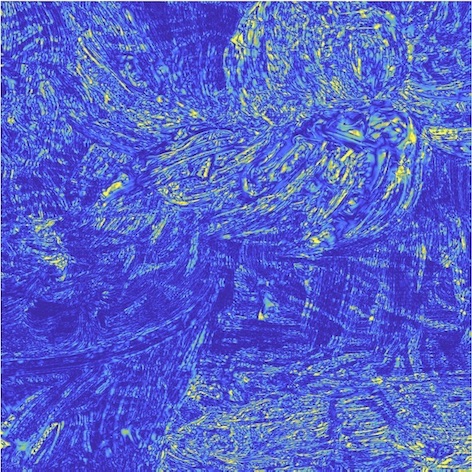}
            \caption{$\lvert I_{180} - I_{0} \rvert$}
        \end{subfigure}%
        \begin{subfigure}[t]{0.23\textwidth}
            \includegraphics[height=1\textwidth]{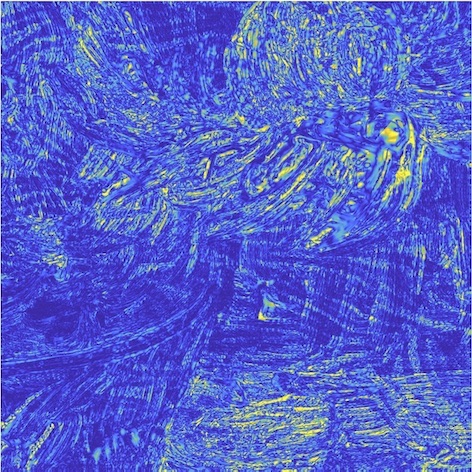}
            \caption{$\lvert I_{270} - I_{0} \rvert$}
        \end{subfigure}%
        \begin{subfigure}[t]{0.08\textwidth}
            \includegraphics[height=3\textwidth]{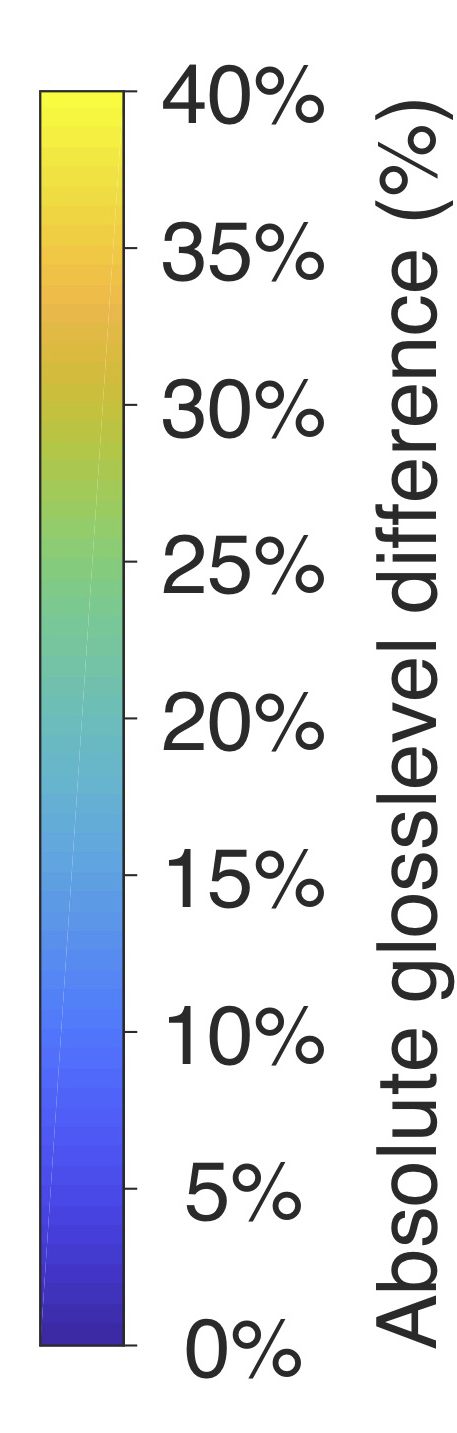}
        \end{subfigure}\\
        \par\bigskip
        \begin{subfigure}[t]{0.23\textwidth}
        \centering
            \includegraphics[height=1\textwidth]{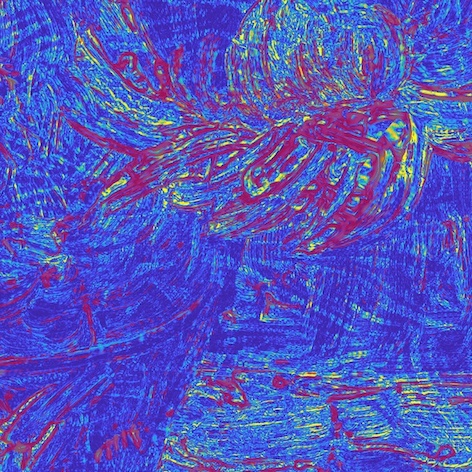}
            \caption{$\lvert I_{90} - I_{0}\rvert$ with combined mask of $I_{90}$ and $I_{0}$ (red)}
        \end{subfigure}
        \hfill%
        \begin{subfigure}[t]{0.345\textwidth}
                \centering
            \includegraphics[width=1\textwidth]{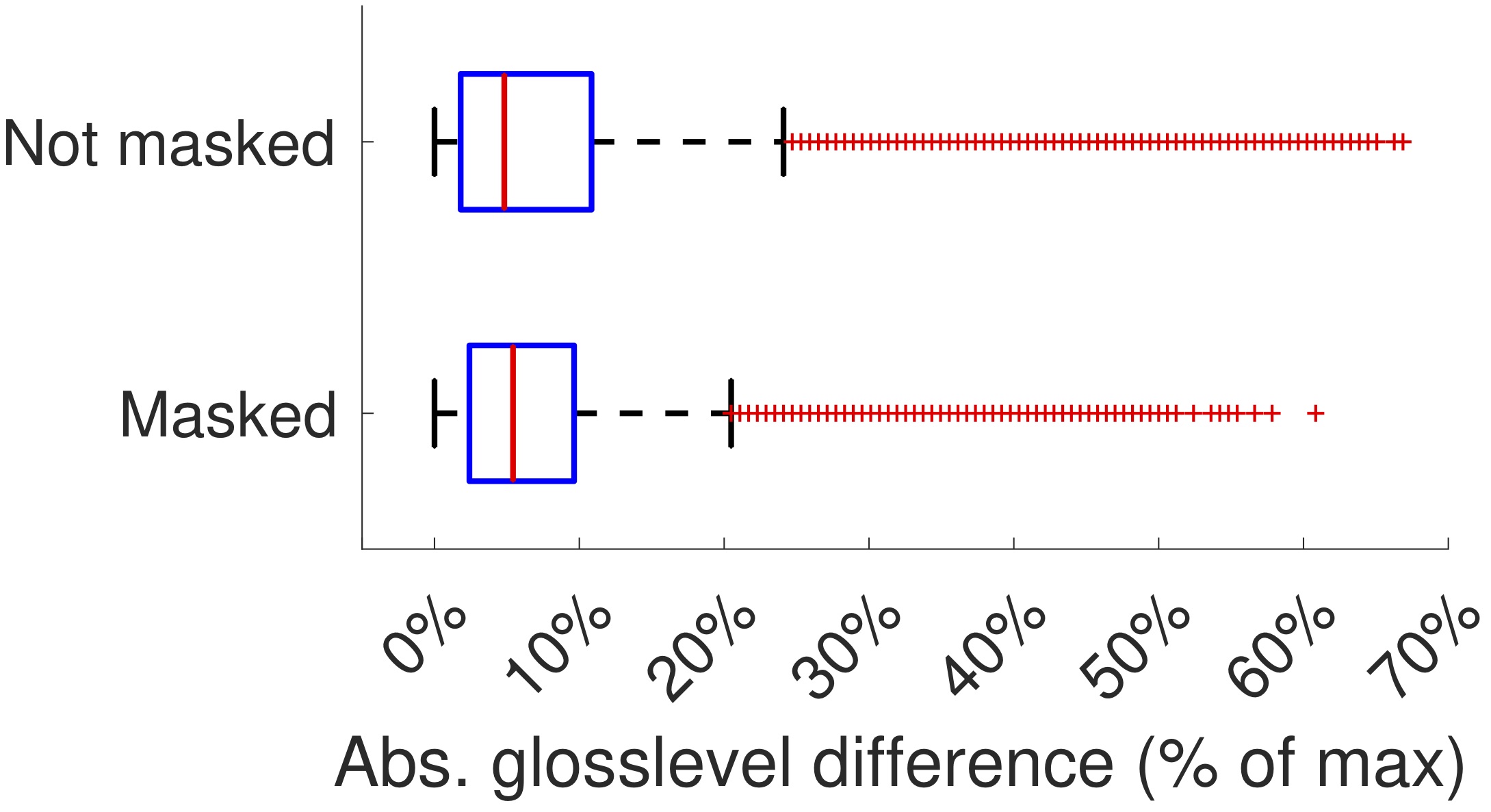}
            \caption{Abs. Glosslevel differences of masked and non-masked areas of ($I_{90} - I_{0}$)}
        \end{subfigure}
        \hfill%
        \begin{subfigure}[t]{0.345\textwidth}
                \centering
            \includegraphics[height=0.667\textwidth]{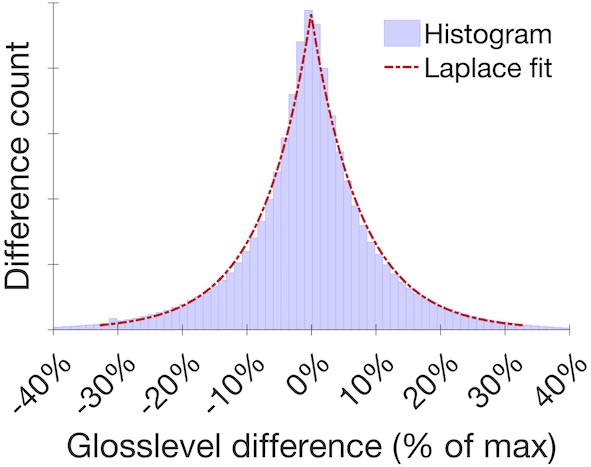}
            \caption{Glosslevel differences of ($I_{90} - I_{0}$)}
        \end{subfigure}
        \hfill\null%
    \caption{Effect of scanning orientation on gloss measurement showing (a) a sample region of the gloss scan $I_{0}$, (b-d) the absolute differences between gloss levels for the scan orientations (see fig. \ref{fig:IlluminationDirections}), displayed between 0 and 40\%, (e) the absolute difference map of ($I_{90} - I_{0}$), with a red, semi-transparent overlay of the combined masks from image $I_{90}$ and $I_{0}$ (see fig. \ref{fig:ResultScan}(d-e)),  
    (f) the boxplots of the absolute differences for the masked and unmasked areas (of $\lvert I_{90} - I_{0} \rvert$, as in (e)),
    (g) shows the histogram of differences (of $I_{90} - I_{0}$), plotted between -40\% and 40\%, as a percentage of the maximum possible difference. All data is scaled between the minimum and maximum measured gloss values (over all images).}
    \label{fig:RotationDifferences}
\end{figure}

\begin{figure}[htb]
\begin{minipage}{.48\textwidth}
\begin{center}
\vspace{0pt}
\captionsetup{type=figure}
\includegraphics[width=0.55\textwidth]{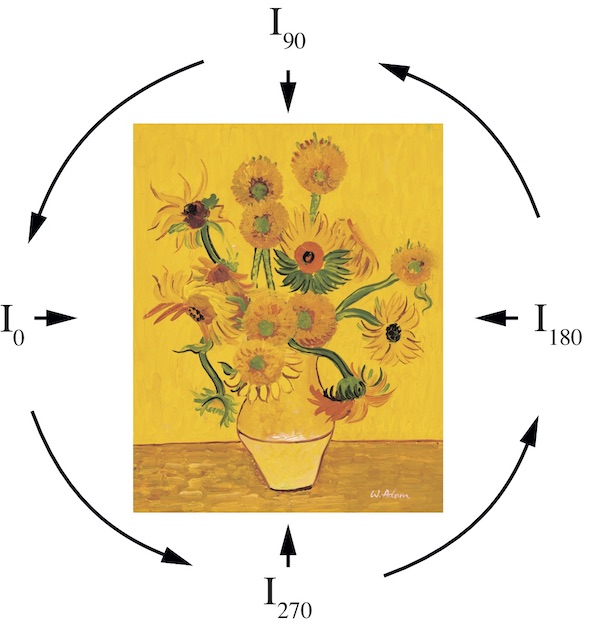}
\captionof{figure}{Illumination directions for the four scans, $I_{0}$, $I_{90}$, $I_{180}$ and $I_{270}$. Note that in practice the painting is rotated \SI{90}{\degree} counterclockwise for every scan} \label{fig:IlluminationDirections}
\end{center}
\end{minipage}%
\hfill
\begin{minipage}{.48\textwidth}
\begin{center}
\vspace{0pt}
\captionsetup{type=table}
\captionof{table}{The table shows the means ($\mu$) and standard deviations ($\sigma$) of the differences between the illuminations directions ($I_0$ to $I_270$) for a sample region (see figure \ref{fig:RotationDifferences}(a)). The data is scaled between the minimum and maximum measured gloss values (over all images)} 
\label{tab:MeanStdHistogram}
\begin{tabular}{lcc}
  \toprule 
  Histogram & $\mu$ & $\sigma$ \\ 
  \midrule
$I_{90} - I_{0}$ & -0.05\% &  10.92\%\\
$I_{180} - I_{0}$ & -0.22\% & 11.69\%\\
$I_{270} - I_{0}$ & -0.21\% & 11.61\%\\
$I_{90} - I_{180}$ & 0.16\% & 12.14\%\\
$I_{270} - I_{180}$ & -0.01\% & 11.13\%\\
$I_{270} - I_{90}$ & -0.17\% & 11.81\%\\
  \bottomrule
\end{tabular} 
\end{center}
\end{minipage}
\end{figure}

To evaluate the effect of the scanning geometry on the measurement, the \textit{Sunflowers} painting was scanned four times, every time rotating the painting by \SI{90}{\degree}. Figure \ref{fig:IlluminationDirections} depicts the effective illumination direction for every scan, labeling the subsequent scans $I_0$ (upright position of the painting), $I_{90}$, $I_{180}$ and $I_{270}$. Note that in reality the painting was rotated relative to the scanner. A sample region of every scan was overlaid, based on local feature matching. The gloss values of the four scans are scaled between the minimum and maximum measured gloss values over all scans. A detail of the four scans and their reciprocal differences are depicted in figure \ref{fig:RotationDifferences}, showing a detail of scan $I_{0}$ in figure \ref{fig:RotationDifferences}(a), and subsequently the absolute differences between $I_{0}$ and the other scanning orientations in figures \ref{fig:RotationDifferences}(b-d). 
(e) shows the absolute difference of $I_{90}$ and $I_{0}$ (as depicted in (b)), with a semi-transparent, red overlay of the combined mask of scan $I_{90}$ and $I_{0}$. (f) shows two boxplots of the absolute differences (for $\lvert I_{90} - I_{0} \rvert$), of the masked and non-masked regions (as shown in (e)). The average differences are respectively $7.0\%$ ($\pm 6.2\%$) and $7.8\%$ ($\pm 8.1\%$) (similar for fig.\ref{fig:RotationDifferences}(c-d)). Figures \ref{fig:RotationDifferences}(e-f) show, that although the masked areas are close to the regions with the largest errors, the infill of the masked regions themselves are not responsible for the largest error. Figure \ref{fig:RotationDifferences}(g) shows the histogram of the differences (for $\lvert I_{90} - I_{0} \rvert$, between -40 and 40\%) with a fitted Laplace distribution, and table \ref{tab:MeanStdHistogram} shows the means and standard deviations of the distributions as depicted in fig. \ref{fig:RotationDifferences}(b-d). The differences are expressed as percentages of the maximum possible difference, meaning that for instance 100\% difference would indicate that $I_{0}$ had a normalized gloss value 0 (completely matte) and $I_{90}$ a normalized gloss value of 100 (high gloss). Note that differences can also occur due to the discrete sampling (at pixel resolution), and the fact that the alignment was not solved on a sub-pixel level, meaning that if you have a high-frequency gloss pattern, you might find differences which are not necessarily the effect of the scanning geometry. Seeing that the differences closely follow a Laplace distribution, we believe that our calibration and off-center corrections are not introducing any systematic error between the images.

\subsection{Print results} \label{subsec:Print results}
\begin{figure} [ht]
\begin{center}
     \setlength{\tabcolsep}{0pt}
\newcolumntype{D}{>{\centering\arraybackslash}p{0.32\linewidth}}
	\begin{tabular}{p{0.4cm} DDD}
\raisebox{0.25\height}{\rotatebox[origin=tl]{90}{{\sf\small No reflections}}} & \multicolumn{3}{l}{\includegraphics[width=0.96\linewidth]{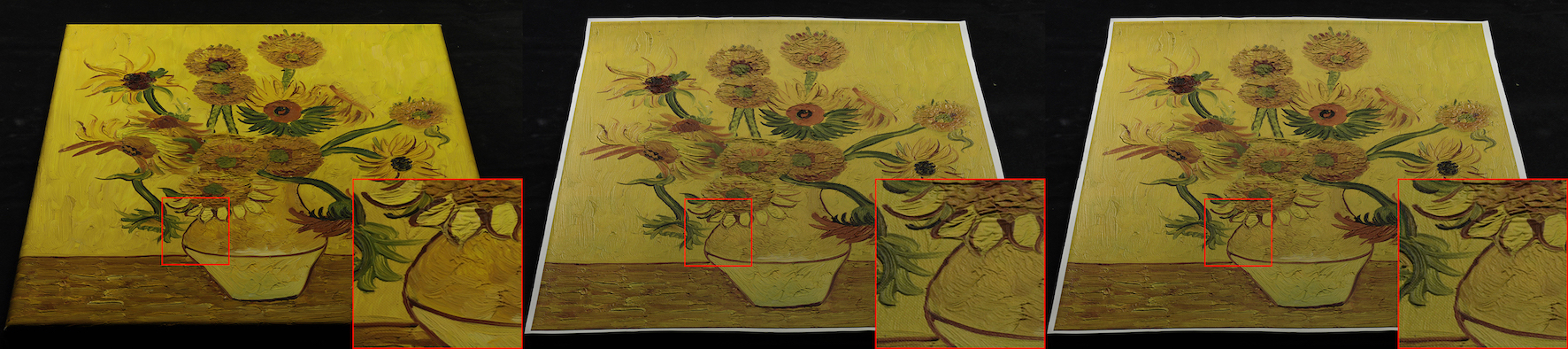}}\\
\raisebox{0.15\height}{\rotatebox[origin=tl]{90}{{\sf\small With reflections }}} & \multicolumn{3}{l}{\includegraphics[width=0.96\linewidth]{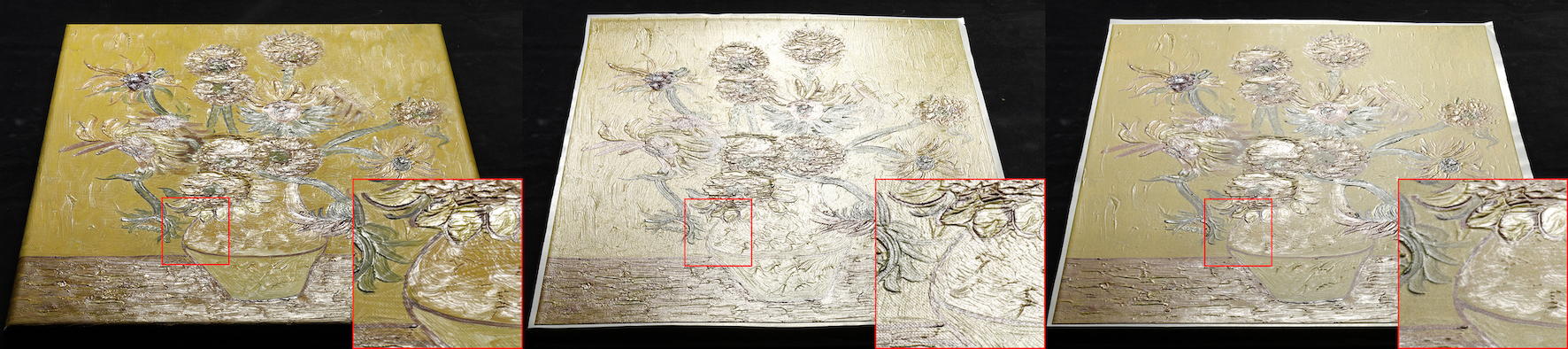}}\\
{\sf\small }&
{\sf\small(a) Painting}&
{\sf\small(b) 3D print, no gloss}&
{\sf\small(c) 3D print, with gloss}\\ 
{\sf\small}&
{\sf\small}& 
{\sf\small modulation layers}&
{\sf\small modulation layers}\\
\end{tabular}
\end{center}
\caption{
Comparison between (a) the painting, (b) a print without gloss modulation layers (which are used to create spatially-varying gloss) and (c) a print with gloss modulation layers, captured under identical mirror angle illumination conditions, without reflections using a polarization filter (top), and including specular reflections (bottom). Note that (b) the print without gloss modulation layers is semi-glossy, which is the intrinsic glossiness of the printed color layers, which exhibits the 'plastic' appearance that participants referred to in previous comparisons to paintings \cite{Elkhuizen2014}.} \label{fig:Print_Results_Total} 
\end{figure}

\begin{figure} [ht]
\begin{center}
     \setlength{\tabcolsep}{0pt}
\newcolumntype{E}{>{\centering\arraybackslash}p{0.24\linewidth}}
	\begin{tabular}{p{0.4cm} EEEE}
  \raisebox{0.2\height}{\rotatebox[origin=tl]{90}{\sf\small 3D \hspace{0.8cm} 2D}} & \multicolumn{4}{l}{\includegraphics[width=0.96\linewidth]{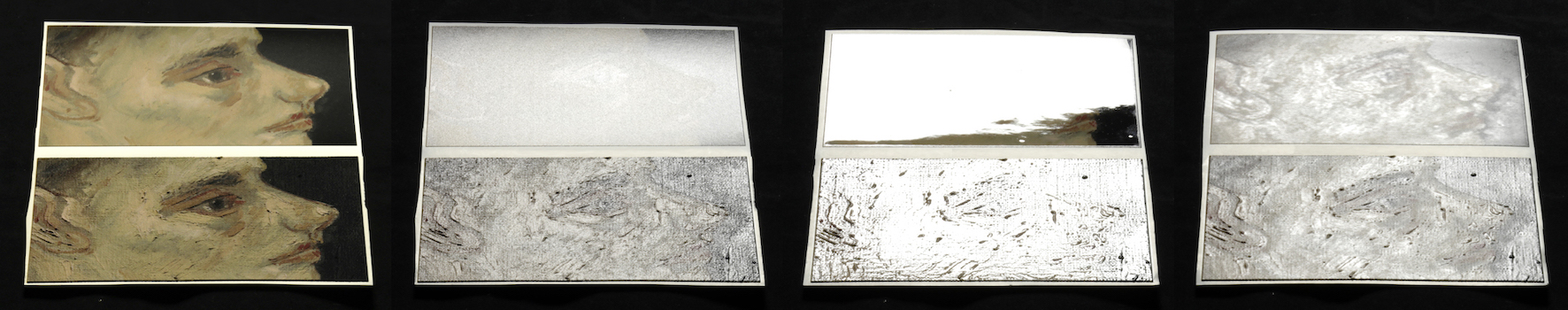}}\\ 
  {\sf\small }&
  {\sf\small (a) No reflections}&
  {\sf\small (b) No gloss modulation layers}&
  {\sf\small (c) First gloss modulation layer}& 
  {\sf\small (d) All gloss modulation layers}\\
\end{tabular}
\end{center}
\caption{Printed details of \textit{Two Wrestling Figures} showing (a) the diffuse color appearance (polarized), (b) the same detail printed without gloss layers, (c) printed with first (high) gloss layer and (d) printed with all gloss layers. The top row has no topography (2D), the bottom row shows the print including topography (3D), both having the same gloss layers. Note that the print without gloss layers is semi-glossy (b), which is the intrinsic glossiness of the color layers.} \label{fig:Print_Results_Details_GvdM}
\end{figure}

\begin{figure} [ht]
\begin{center}
     \setlength{\tabcolsep}{0pt}
\newcolumntype{E}{>{\centering\arraybackslash}p{0.24\linewidth}}
	\begin{tabular}{p{0.4cm} EEEE}
  \raisebox{0.2\height}{\rotatebox[origin=tl]{90}{\sf\small 3D \hspace{0.8cm} 2D}} & \multicolumn{4}{l}{\includegraphics[width=0.96\linewidth]{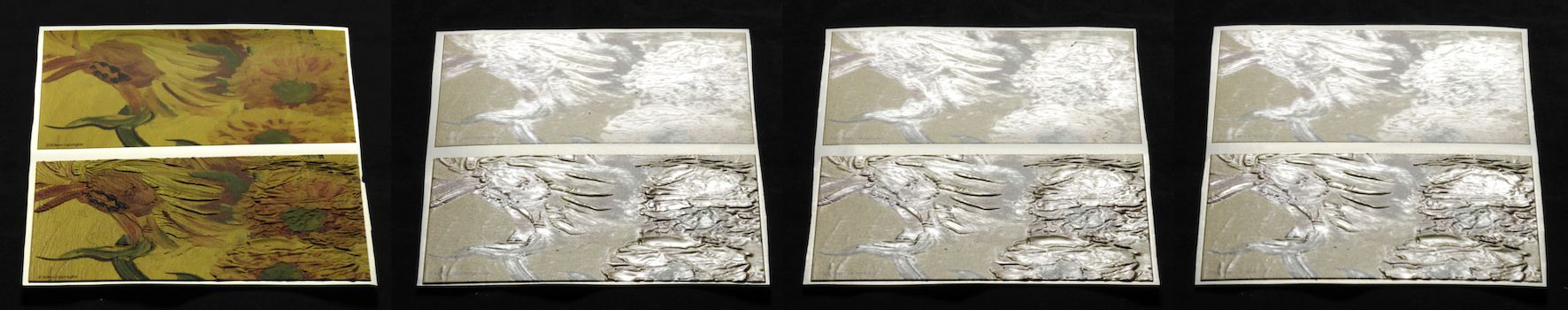}}\\
  {\sf\small }& 
  {\sf\small (a) No reflections}&
  {\sf\small (b) Glossmap $I_{90}$}&
  {\sf\small (c) Glossmap $I_{180}$}&
  {\sf\small (d) Glossmap $I_{270}$}\\
\end{tabular}
\end{center}
\caption{Printed details of \textit{Sunflowers} showing (a) the diffuse color appearance (polarized), (b) the same detail of glossmap $I_{90}$, (c) glossmap $I_{180}$, and (d) glossmap $I_{270}$. The top row has no topography (2D), the bottom row shows the print including topography (3D), both having the same gloss modulation layers.} \label{fig:Print_Results_Details_SF}
\end{figure}

Printing the reproductions takes 1/2 hour for the \textit{Sunflowers} painting and 3 hours for the \textit{Two Wrestling Figures} painting.
Figure \ref{fig:Print_Results_Total} shows the \textit{Sunflower} painting and 3D printed reproduction, photographed under an identical mirror illumination and viewing angle.
Figure \ref{fig:Print_Results_Total}(a) depicts the painting, \ref{fig:Print_Results_Total}(b) the print without gloss layers added (showing the default, uniform gloss appearance of Oc\'{e}'s Elevated Printing process), and \ref{fig:Print_Results_Total}(c) showing the print with spatially-varying gloss. In the top row of the figure, the reflections are removed using a polarization filter, to better visualize the diffuse color appearance; the bottom row shows the images with the specular reflections included, which is most like their appearance to the human eye, when viewing them at this angle. These results show that the scan data can successfully be used to reproduce the spatially-varying gloss of the painting's surface, although it is apparent that the most matte regions on the painting (fig. \ref{fig:Print_Results_Total}(a)), appear more matte than the most matte regions in the reproduction (Fig.\ref{fig:Print_Results_Total}(b)). This is most clear in the yellow background, where the reproduction appears glossier than the painting, as the printer is unable to match the matteness level of the painting.

Figure \ref{fig:Print_Results_Details_GvdM} shows a detail of the \textit{Two westling figures} prints, where figure \ref{fig:Print_Results_Details_GvdM}(a) shows the diffuse color appearance (polarized), figure \ref{fig:Print_Results_Details_GvdM}(b) shows no gloss layers printed, figure \ref{fig:Print_Results_Details_GvdM}(c) shows the first high gloss layer printed, and \ref{fig:Print_Results_Details_GvdM}(d) shows the print when all gloss layers are added. The top part is printed \emph{without} elevation and the bottom part \emph{with} elevation, to emphasize the contribution of the printed gloss variation on the appearance at this viewing angle. 

Figure \ref{fig:Print_Results_Details_SF} shows a printed detail of the \textit{Sunflowers} where \ref{fig:Print_Results_Details_SF}(a) shows the diffuse color appearance (polarized), \ref{fig:Print_Results_Details_SF}(b) the same area captured when the painting is rotated by \SI{90}{\degree} ($I_{90}$), \ref{fig:Print_Results_Details_SF}(c) when rotated by \SI{180}{\degree} ($I_{180}$), and \ref{fig:Print_Results_Details_SF}(d) when rotated by \SI{270}{\degree} ($I_{270}$). Results show  visually very similar spatially-varying gloss characteristics, thereby suggesting that the scan orientation is of limited influence on the reproduced gloss. 

\section{Discussion} \label{sec:Discussion}
The proposed system demonstrates that it is possible to capture spatially-varying gloss of a painting's surface with sufficient accuracy for the purpose of 3D printing. Visual inspection indicates that the approach is able to authentically reproduce the color, topography and spatially-varying gloss of a painting.  The approach seems effective in trading angular measurement accuracy (present in many BRDF capturing approaches for rendering purposes) for speed and spatial resolution, making it suitable for 3D printed appearance reproduction. Additionally, we are able to print a range of gloss levels with one transparent material - in analogy to CMYK(W) printing, which creates a wide range of colors with only 5 materials - making it a suitable starting point for further full appearance (re)production workflows. 

\subsection{Scanning paintings and (other) heavily textured surfaces}
The current setup is configured for materials with a refractive index typical for (varnished) oil paintings. Therefore the scanner will need to be reconfigured if a surface with a very different refractive index needs to be scanned. Moreover, a rough estimate of the refractive index is needed, and therefore some knowledge of the surface material, in order to configure the scanner.

A limitation of the current approach is the need for off-center normalization of the scan measurements, whereby the further the measurement is away from the exact Brewster's angle at center of the image, the more correction is needed. In other words, there is a trade-off between speed (larger scan area) and accuracy of the measurement. Another limitation is that with heavily textured surfaces, a large portion of the gloss map is masked and has to be filled by interpolation, leading to less accurate results. This approach will therefore work best on surfaces which are relatively flat, which is generally the case for pre-modern panel and canvas paintings. We would like to argue that paintings with moderate height variations (like the ones shown in this paper) are also reproducible. However, paintings were the paint encloses empty space between the paint and the canvas (creating 'overhangs' in 3D printing terminology) cannot be reproduced by the proposed 3D scanning approach, nor the 3D printing system, due to the method of plane projection, and the lack of a removable support material.

\subsection{Influence of scanning geometry on gloss measurements}
The results also show that there is a consistency between the scan results from the four rotation angles ($I_{0}$ - $I_{270}$), meaning the influence of the scanning geometry on the measurement is limited. Here we should note that some of the differences found (as presented in figure \ref{fig:RotationDifferences}) might be explained by image misalignment and the discrete sampling of high-frequency gloss variation. 

Figure \ref{fig:RotationDifferences}(e) shows that largest errors are not found directly at the masked regions, indicating that the masking and infill itself is not causing the largest error. However, the areas with the largest errors are found in the vicinity of the masked areas. From theory it can be expected that the largest errors in gloss measurement will occur in \emph{high} gloss regions, where the surface normal varies (due to height variations). This is due to the increased concentration of the specular reflection and around the mirror reflection angle for high gloss surfaces, whereby only a small deviation in surface normal, causes a large difference in measurement. In our sample area (fig. \ref{fig:RotationDifferences}(a), coincidently the high gloss regions are also the regions with larger height variation in our sample. We can therefore not directly disentangle this effect based on scan of these paintings, which are in essence under defined. No significant difference was found in the shape of the histogram if we compare a high gloss area to a low gloss area, within our sample.

\subsection{Other interactions between gloss, color, and topography scanning and fabrication}
Although the gloss map seem to depict the spatial variation and intensity of gloss, proportionally to the visual sensation, it remains to be investigated if the gloss measurement is completely independent of the diffuse color. Additionally, it remains to be investigated what the effect is in terms of scanning as well as fabrication of the gloss on the height. Currently it is unknown what the effect of (semi)-transparent varnish layers is on the height measurement. Likewise, the effect of the printed gloss layers has an effect of the surface topography of the print. The extent of this remains to be investigated, as this is currently not corrected for.

\subsection{Scanning speed and capturing larger paintings}
Whereas we argue that the current scanning and image processing times (as mentioned in section \ref{subsec:Scan results}) are reasonable for the purposes of conducting a case study, the authors would like to emphasize that the current implementation was not optimized for scanning and processing speed. If the capturing procedure was to be optimized for speed, the exposure time of the images, and overhead time for movement and storing data are of main importance. Assuming a (realistic) exposure time of 2 seconds per image (and overhead of 2 seconds), it would be possible to capture \SI{1}{\metre\squared} within 1 hour. Capturing larger areas would require either creating a larger frame (which can be done up to a certain extent), or repositioning the frame. The capturing method (carried out by the equipment on the scanner platform) would remain the same. The printable area is currently limited to \SI{1.25x2.5}{\metre}, but prints might also be tiled to create larger areas. The printable area is not easily extended.

\subsection{Gloss mapping and perceptual evaluation}
In terms of fabrication, the measured gloss values are mapped to the full range of printable gloss levels. Scanning printed reference samples, may discover better relations between the measured gloss values and the printable gloss levels. In the (likely) case that the printable range of gloss is not sufficient, a strategy should be developed for gloss mapping - in analogy to gamut mapping of color - through psychophysical experiments. 

Additionally, further quantitative and qualitative evaluation of the reproductions should be conducted to validate the scanning and printing results, now that the printed results are at a level suitable for perceptual testing. Finally, we would like to argue that, when viewing figure \ref{fig:Print_Results_Details_GvdM} and \ref{fig:Print_Results_Details_SF}, we can conclude that reproducing spatially-varying gloss provides added perceptual value, only when it is reproduced in conjunction with other material appearance attributes, like color and topography. Reproducing the spatially-varying gloss and color, without the topography, does not bring the appearance closer to the original than visa versa. 

\section{Conclusions} \label{sec:Conclusions}
In this paper, we present a painting appearance reproduction system with a focus on capturing gloss appearance. The spatially-varying gloss of a painting is measured by sampling the specular reflection close to Brewster's angle. A mathematical model is developed to normalize off-center deviations of the gloss measurements. Deviations in the local surface normal, as well as shadows are masked by the height map and filled with relevant gloss information. Experiment results indicate that the proposed system is able to simultaneously reproduce the color, the 3D topography and especially the gloss information of a painting. Figure \ref{fig:Print_Results_Total} shows how the simultaneous reproduction of color, topography and gloss variations results in a visually convincing 3D print. Figure \ref{fig:Print_Results_Details_SF} shows that the effects of the scanning orientation are limited with respect printed results.

Limitations of the system are also identified regarding the off-center normalization, local deviations in surface normals and shadowing, possible color dependency, and relations between gloss measurements and printable gloss levels, which highlight the future work of the authors.

\begin{acks}\label{sec:Acknowledgement}
The authors would like to express their appreciation to the team of `Het Geheim van de Meester' for providing the painting \textit{Two Wrestling Figures} in the style of Van Gogh which was used as one of the case studies in this paper. This research, as part of the \textit{3D Fine Art Reproduction Project} was funded by Oc\'{e} Technologies B.V.
\end{acks}
 
\bibliographystyle{ACM-Reference-Format}
\bibliography{JOCCH_paper_-_gloss_reproduction.bib} 

\end{document}